\documentclass[12pt]{article}
\bigskip

\usepackage{amsmath}
\usepackage{amsfonts}
\usepackage{amssymb}
\usepackage[bf]{caption}
\usepackage{cite}
\usepackage{color}
\usepackage{psfrag,graphicx}
\usepackage{shadow}
\usepackage{subfigure}
\usepackage{textcomp}
\usepackage{wrapfig}
\usepackage{upgreek}
\usepackage{rotating}

\newcommand{\cte}[1]{``#1"}

\newcommand{\fetlut}[1]{\mbox{\boldmath{\mbox{$#1$}}}} 

\newcommand{\dirac}[1]{\updelta\!\left({#1}\right)}

\newcommand{\ep}{\varepsilon}

\newcommand{\ds}{\displaystyle}

\renewcommand{\j}{\text{j}}

\newcommand{\w}{\omega}
\newcommand{\jw}{\j\omega}
\newcommand{\e}{\text{e}}

\newcommand{\br}[1]{\left({#1}\right)}

\newcommand{\brn}[1]{\!\left({#1}\right)} 

\newcommand{\brh}[1]{\left[{#1}\right]}
\newcommand{\brc}[1]{\left\{{#1}\right\}}
\newcommand{\brb}[1]{\left|{#1}\right|}

\newcommand{\lap}{\nabla^2}
\newcommand{\lapT}{\nabla_{\text{t}}^2}
\newcommand{\rot}{\nabla\times}

\newcommand{\grad}{\nabla}
\newcommand{\gradT}{\nabla_{\text{t}}}

\newcommand{\divT}{\nabla_{\text{t}}\cdot}

\newcommand{\kt}{k_\text{t}}

\newcommand{\pdtv}[2]{\ds\frac{\partial^2 #1}{\partial {#2}^2}}

\newcommand{\vc}[1]{\fetlut{#1}}   
\newcommand{\vs}[1]{\vc{\scriptstyle #1}}   

\newcommand{\uv}[1]{\hat{\vc{#1}}} 
\newcommand{\un}{\uv{n}}

\newcommand{\ux}{\uv{x}}
\newcommand{\uy}{\uv{y}}
\newcommand{\uz}{\uv{z}}

\newcommand{\rhov}{\vc{\rho}}

\newcommand{\kvt}{\kvv_\text{t}}
\newcommand{\kvst}{\vs{k}_\text{t}}

\newcommand{\Evv}{\vc{E}}

\newcommand{\Hvv}{\vc{H}}

\newcommand{\Jvv}{\vc{J}}
\newcommand{\Kvv}{\vc{K}}

\newcommand{\evv}{\vc{e}}
\newcommand{\fvv}{\vc{f}}
\newcommand{\gvv}{\vc{g}}

\newcommand{\kvv}{\vc{k}}

\newcommand{\pvv}{\vc{p}}

\newcommand{\rvv}{\vc{r}}

\DeclareMathAlphabet{\mathsfbfsl}{T1}{cmss}{bx}{sl}

\newcommand{\dd}{\mbox{d}}
\newcommand{\dV}{\dd v}
\newcommand{\dS}{\dd s}

\newcommand{\dA}{\dd a}
\newcommand{\dl}{\dd l}

\newcommand{\bJ}[2]{\mbox{J}_{#1}\!\br{#2}}

\newtheorem{figtext}{Figure}

\newtheorem{tabtext}{Table}

\begin{document}

\title{{\Large \bf On the spectral domain approach to long-range
propagation of high-frequency waves along a strip conductor above a PEC
surface}}


\author{Martin Norgren}
\maketitle

\begin{center}
Electromagnetic Engineering Lab\\ Royal Institute of
Technology\\ SE-100 44 Stockholm, Sweden \\[6mm]
Corresponding author: Martin Norgren\\
Email: martin.norgren@ee.kth.se\\ Tel: +46 8 7907410; Fax: +46 8
205268\\[10mm]
\end{center}

\newpage

\begin{abstract}

 A generic problem of high frequency wave propagation along a metallic strip in parallel above a PEC ground plane is considered. The wave is excited by an elemental electric dipole at an arbitrary location above the PEC plane. The full wave problem, for arbitrary widths of the strip, is solved by means of a mode matching approach and expansion of the strip surface current into Chebyshev polynomials. For narrow strips, an approximate method using only longitudinal currents is derived, and compared numerically with the full wave method. Utilizing the concept of equivalent radius, the approximate method for narrow strips is evaluated numerically against results for thin circular wires. It is concluded that the approximate method is suitable for handling multiple wires in layered structures, wherefore the method has potential usefulness for estimating long range propagation of high frequency waves in wire structures like power lines and railway feeding systems, containing over-head wires and wires submerged into ground.

\end{abstract}

\section{Introduction}

The analysis of electromagnetic waves propagating along wires
parallell with interfaces, separating regions having different
electromagnetic properties, is of importance when considering
electrical systems, like power lines \cite{LazaropoulosCottis},
submerged cables \cite{PoljakETAL2,TheethayiETAL2} and the power
supply for railways \cite{CozzaDemoulin}. In such power supply
systems, the waves can be unintentional, triggered by internal
and/or external disturbances
\cite{CozzaDemoulin,PetracheETAL,SeeETAL}, or intentional, for e.g.
monitoring and communication purposes
\cite{SartenaerDelogne,AmirshahiKavedrah,LazaropoulosCottis}.

The literature on methods for analyzing parallell conductor
structures is vast, wherefore we will not give a comprehensive
survey here; a selection of references can be found in the monograph
\cite{RachidiTkachenko}. In a low frequency application, it is
suitable to model the structure as a multiconductor transmission
line (MTL), and solve the propagation problem by means of the
quasi-TEM mode theory \cite{Lindell1981,LindellGu1987}, generalized
to account for the coupling to external sources
\cite{RachidiTkachenko}(chapter 1). The basic assumption is that at
the highest frequency, the wavelength is large compared with the
transverse dimensions of the system. For power lines and railway
feeding systems one can thus reliably use MTL theory at frequencies
below about 1-10 MHz. For frequencies at which the wavelength
becomes comparable with the wire separations the presence of higher
order radiating modes cannot be neglected. In such cases one can use
full wave, or nearly full wave, methods, like generalized MTL
methods \cite{RachidiTkachenko}(chapter 4) or other methods, like
the Finite Difference Time-Domain method
\cite{TheethayiETAL1} and integral equation based methods
\cite{PoljakETAL1,AlyonesETAL}.

Typically, the full wave methods are based on numerical
discretization of wires of finite length, whereby for very long
wires the required numerical accuracy may yield equations that are
too massive from the computational point of view. In particular,
this will be the case if one wants to study electromagnetic
interference (EMI) and communication at frequencies of orders from
100 MHz to several GHz, where a typical power line becomes
electrically very large also in the transversal directions. Hence,
in order to estimate e.g. possible distances for GHz communication
along wire structures and EMI effects at the same frequencies one
needs methods capable of handling very long (in terms of
wavelengths) wire structures.

One example of a nearly full wave method for an infinitely long
uniform MTL is presented in \cite{BridgesShafai}. The MTL is located
above a lossy halfspace and illuminated by a plane wave with an
arbitrary direction of incidence; remotely located dipole sources
are also considered. The uniformity facilitates the use of a spatial
Fourier transform whereby the problem can be solved in the transform
space, i.e. the spectral domain. Except for the thin wire
approximation, this method is exact.

Spectral domain methods have been used frequently for studying
propagation along MTL structures in the form of thin planar strip
conductors on substrates, see e.g. Sect. 4.8 in \cite{Collin2} for a calculation
of the transmission line parameters for quasi-TEM modes propagating
along a microstrip line. Hence, it is of interest to
investigate whether these methods can be useful also for the above
mentioned power line problems. A motivation for considering methods
developed for planar conductors is that if the thin wire
approximation holds, and proximity effects can be neglected, a wire
of quite arbitrary cross-section can be replaced with an equivalent
strip conductor. For example, for a circular wire
the equivalent strip has a width that is twice the diameter of the wire \cite{Norgren2003}. In
\cite{Butler}, the two-dimensional scattering problem for a strip
illuminated by a plane wave has been solved, for both polarizations
of the incident electric field. Excitation/scattering problems for
strip gratings using localized nearby sources have been considered
in \cite{LovatETAL,KaganovskyShavit}. In the context of
power/railway lines, localized sources are of relevance when
studying EMI emanating from e.g. spark-overs in insulators or from
pantograph arcing \cite{Midya}. Another situation where localized
sources are of relevance is high frequency communication, where the
guided wave on the wire structure can be excited by coupling to a
suitably placed antenna.

In this paper, we consider a metallic strip placed in parallel above
a perfectly electrically conducting (PEC) plane. The configuration
is excited by an electric dipole at an arbitrary location above
the PEC plane. For the analysis, we use a spectral domain method,
based on Fourier transforms in both of the horisontal directions.
The analysis is systemized in terms of classical waveguide theory,
where the spectral components are treated as waveguide modes in the
vertical direction \cite{FelsenMarkuwitz1994}(Section 3.2b). Hence,
the strip can be considered as a generalized diaphragm in a plane
connecting two waveguide sections; a similar point of view has also
been used in \cite{LovatETAL}. Like in \cite{Butler}, we will in the
context of the strip model make no approximations, whereby the
analysis holds for arbitrary widths of the strip, in terms of
wavelengths. Furthermore, there are no restrictions on the
separation distance between the wire and the PEC plane.

The paper is organized as follows. In Section
\ref{section:Problem_formulation_and_theory}, we formulate the
problem and derive the spectral method. For clarity, the derivation
is carried out in some detail, with additional details in the
appendices. The result is a linear system for determining the
coefficients of the spectral surface current density on the strip.
In Section \ref{section:Numerical_results}, we present numerical
results. First, we consider a wide strip, whereby we have to use the
exact (truly full wave) method. Then we consider a narrow strip and
investigate the applicability of certain approximations for that
case. Finally, we evaluate the equivalent radius concept, by
comparing the results for narrow strips with results for a circular
wire: analytical results for the TEM-mode approximation and full-wave results obtained by the MoM based Numerical Electromagnetics Code
(NEC) \cite{NEC}. The method and the results are summarized and
discussed in Section \ref{section:Discussion_and_conclusions}.

\section{Problem formulation and theory}

\label{section:Problem_formulation_and_theory}

\begin{figure}[h]
\centering \psfrag{1}{$y_0$} \psfrag{m}{$-h$} \psfrag{p}{$h$}
\psfrag{x}{$x$} \psfrag{y}{$y$} \psfrag{z}{$z$} \psfrag{0}{$z_0$}
\psfrag{a}{$z=a$} \psfrag{1}{$y_0$} \psfrag{2}{$x_0$}
\psfrag{u}{$\pvv$}
\includegraphics[scale=1.1]{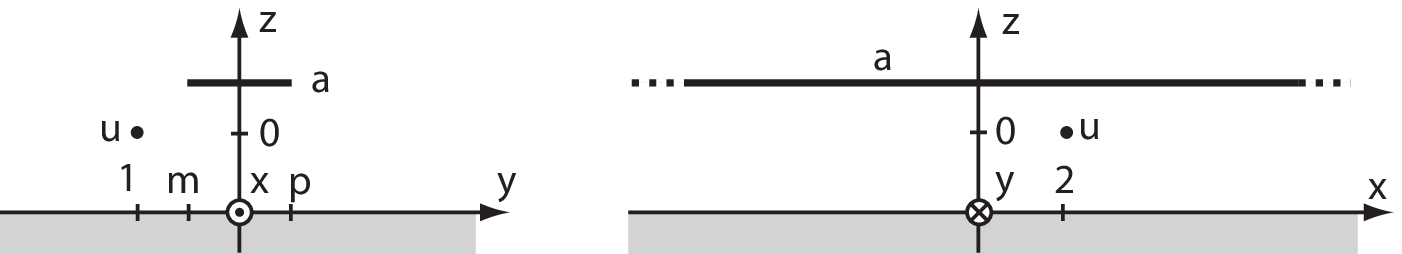}
\caption{ } \label{Fig:1}
\end{figure}

The problem geometry is depicted in Figure \ref{Fig:1}. There is a
PEC surface in the plane $z=0$. In the plane $z=a$ there is an
infinitely thin PEC strip that is infinite in the $x$-direction and
with edges at $y=\pm h$. At the location $\rvv_0=x_0\ux + y_0\uz +
z_0\uz$ there is an electric dipole source with dipole moment
$\pvv$. The medium in the region $z>0$ is homogeneous, isotropic and
in general lossy, described by the complex permittivity $\ep$ and
the complex permeability $\mu$, which are in general frequency
dependent.

The problem is to find the fields and the strip surface current that
results from to the radiation from the dipole.


\subsection{Preliminaries}

We start from the Maxwell equations in a homogeneous isotropic
medium:
\begin{align}
\rot\Evv\brn{\rvv} & = -\j k \eta\Hvv\brn{\rvv}, \label{eq:prelim:Maxwell1}\\
\eta\rot\Hvv\brn{\rvv} & = +\j k \Evv\brn{\rvv} +
\eta\Jvv\brn{\rvv}, \label{eq:prelim:Maxwell2}
\end{align}
where $\Evv$ and $\Hvv$ are the electric and magnetic fields,
respectively, $k=\w\sqrt{\ep\mu}$ is the medium wave number,
$\eta=\sqrt{\mu/\ep}$ is the medium wave impedance and $\Jvv$ is the
source current density. $\w$ is the angular frequency, in a
suppressed $\e^{\jw t}$ time-dependence.

We decompose the radius vector into $\rvv  = \rhov + z\uz $, where
the transversal part $\rhov = x\ux + y\uy$, and introduce the
following pair of Fourier-transforms:
\begin{align} & \Evv\brn{\kvt, z}  = \int_\mathcal{S}
\Evv\brn{\rvv}\e^{\j
\kvst\cdot\vs{\rho}} \dS, \label{eq:prelim:FT1}\\
& \Evv\brn{\rvv}  = \frac{1}{4\pi^2}\int_\mathcal{K} \Evv\brn{\kvt,
z}\e^{-\j \kvst\cdot\vs{\rho}} \dd^2 \kt, \label{eq:prelim:FT2}
\end{align}
defined analogously for $\Hvv, \Jvv$ etc.. In (\ref{eq:prelim:FT1}),
$\dS = \dd x \dd y$ and $\mathcal{S}$ denotes the entire $xy$-plane.
The spectral (Fourier) variable $\kvt = k_x\ux + k_y \uy$ is a
transverse wave-vector. In (\ref{eq:prelim:FT2}), $\dd^2\kt = \dd
k_x \dd k_y$ and $\mathcal{K}$ denotes the entire  $k_x k_y$-plane.
In the following, we refer to $\Evv\brn{\rvv}$ as the spatial
electric field and to $\Evv\brn{\kvt, z}$ as the spectral electric
field, and similarly for other quantities.

\subsection{TM- \& TE-modes}

In any slab region, $z_1 < z < z_2$, where $\Jvv\brn{\rvv}=\vc{0}$,
it follows from (\ref{eq:prelim:Maxwell1}) and
(\ref{eq:prelim:Maxwell2}) that the fields satisfy the homogeneous
Helmholtz equation
\begin{align}
\br{\lap + k^2} \begin{bmatrix}
                        \Evv\brn{\rvv} \\
                        \Hvv\brn{\rvv}
                      \end{bmatrix}
 = \begin{bmatrix}  \vc{0} \\ \vc{0}\end{bmatrix}
 \label{eq:modes:Helmholtz1}
\end{align}
Using (\ref{eq:prelim:FT2}) to express $\Evv\brn{\rvv}$ and
$\Hvv\brn{\rvv}$ in (\ref{eq:modes:Helmholtz1}), and using the
property $\grad\e^{-\j\kvst\cdot\vs{\rho}}= -\j\kvt
\e^{-\j\kvst\cdot\vs{\rho}}$, it follows that the spectral fields
satisfy
\begin{align}
\br{-\kt^2 + \pdtv{}{z} + k^2} \begin{bmatrix}
                        \Evv\brn{\kvt, z} \\
                        \Hvv\brn{\kvt, z}
                      \end{bmatrix}
 = \begin{bmatrix}  \vc{0} \\ \vc{0}\end{bmatrix},
 \label{eq:modes:Helmholtz2}
\end{align}
with solutions of the form
\begin{align}
\Evv\brn{\kvt, z}, \Hvv\brn{\kvt, z} \propto \e^{\mp\j k_z z}
\end{align}
where the longitudinal wavenumber
\begin{align}
k_z = \sqrt{k^2 - \kt^2} \label{eq:prelim:kz}
\end{align}
Since any physical medium exhibits losses, it follows from the
passivity conditions \cite{LindellETAL1994} that $\text{Im}\brc{\ep}
< 0$ and that $\text{Im}\brc{\mu} < 0$, resulting in that $k$ and
$k_z$ both are complex. The branch of the square root in
(\ref{eq:prelim:kz}) is always chosen such that $\text{Im}\brc{k_z}
< 0$. Hence, $\e^{-\j k_z z}$ represents a solution decaying in the
$+z$-direction, while $\e^{+\j k_z z}$ represents a solution
decaying in the $-z$-direction.

Using (\ref{eq:prelim:FT2}) to express the fields in Maxwell's
equations (\ref{eq:prelim:Maxwell1}) and (\ref{eq:prelim:Maxwell2}),
with $\Jvv= \vc{0}$ and $z$-dependencies of the forms $\e^{\mp\j k_z
z}$, it is shown in Appendix \ref{section:app:transverse} that the
transversal spectral field components can be expressed in terms of
the longitudinal components:
\begin{align}
\Evv_\text{t}^\pm & = \frac{-1}{\kt^2} \brh{ \pm k_z \kvt E_z^\pm -
k
\uz\times\kvt \eta H_z^\pm}, \label{eq:modes:Et}\\
\eta \Hvv_\text{t}^\pm & = \frac{-1}{\kt^2} \brh{ \pm k_z \kvt \eta
H_z^\pm + k \uz\times\kvt E_z^\pm}, \label{eq:modes:Ht}
\end{align}
where $\pm$ refers to the $\pm z$-decaying solutions.

From the structures of (\ref{eq:modes:Et}) and (\ref{eq:modes:Ht}),
we find it convenient to introduce the dimensionless functions
\begin{align}
\fvv\brn{\kvv_\text{t}} & = \frac{k_z}{k_\text{t}^2}\kvv_\text{t},
\label{eq:modes:f}
\\
\fvv^\pm\brn{\kvv_\text{t}} & = \fvv\brn{\kvv_\text{t}} \mp \uz,
\label{eq:modes:fpm}
\\
\gvv\brn{\kvv_\text{t}} & = \frac{k}{k_\text{t}^2}
\uz\times\kvv_\text{t}, \label{eq:modes:g}
\end{align}
subject to
\begin{align}
& \fvv^\pm\brn{\kvt}\cdot\gvv\brn{\kvt}=\fvv\brn{\kvt}\cdot\gvv\brn{\kvt}=0. \label{eq:modes:fgort}
\end{align}
Hence, the transverse magnetic (TM) parts of the spectral fields are
written
\begin{align}
\Evv^\text{TM}\brn{\kvt, z} & = A^+\brn{\kvt}\fvv^+\brn{\kvt}\e^{-\j
k_z z} + A^-\brn{\kvt}\fvv^-\brn{\kvt}\e^{+\j k_z z} \label{eq:modes:TM_E}\\
\eta\Hvv^\text{TM}\brn{\kvt, z} & = \gvv\brn{\kvt}
\brh{A^+\brn{\kvt}\e^{-\j k_z z} - A^-\brn{\kvt}\e^{+\j k_z z} }
\label{eq:modes:TM_H}
\end{align}
while the transverse electric (TE) parts of the spectral fields are
written
\begin{align}
\Evv^\text{TE}\brn{\kvt, z} & = \gvv\brn{\kvt}
\brh{B^+\brn{\kvt}\e^{-\j k_z z} + B^-\brn{\kvt}\e^{+\j k_z z} } \label{eq:modes:TE_E}\\
\eta\Hvv^\text{TE}\brn{\kvt, z} & =
-B^+\brn{\kvt}\fvv^+\brn{\kvt}\e^{-\j k_z z} +
B^-\brn{\kvt}\fvv^-\brn{\kvt}\e^{+\j k_z z} \label{eq:modes:TE_H}
\end{align}
For each value of $\kvt$ there is one TM-mode and one
TE-mode\footnote{When $\kvt=\vc{0}$ there are no longitudinal
fields, and instead two degenerate TEM-modes (with $k_z = k$).}.
$A^\pm\brn{\kvt}$ and $B^\pm\brn{\kvt}$ are mode coefficients, which
are determined by the sources and the boundary conditions. Once the
mode coefficients have been determined, the spatial fields are
determined using (\ref{eq:prelim:FT2}), resulting in
\begin{align} \Evv\brn{\rvv} & = \frac{1}{4\pi^2}\int_\mathcal{K}
\brh{\Evv^\text{TM}\brn{\kvt, z} + \Evv^\text{TE}\brn{\kvt,
z}}\e^{-\j \kvst\cdot\vs{\rho}} \dd^2 \kt,
\label{eq:modes:E_spatial} \\
\Hvv\brn{\rvv} & = \frac{1}{4\pi^2}\int_\mathcal{K}
\brh{\Hvv^\text{TM}\brn{\kvt, z} + \Hvv^\text{TE}\brn{\kvt,
z}}\e^{-\j \kvst\cdot\vs{\rho}} \dd^2 \kt.
\label{eq:modes:H_spatial}
\end{align}

\subsection{The spectral fields from the dipole, the ground plane and the strip}

In Appendix \ref{section:app:excitation} we show that the spectral
fields from the dipole become
\begin{align}
\Evv^{(\text{d})}\brn{\kvv_\text{t}, z} & = A_1\brn{\kvv_\text{t},
z}\brh{\text{H}\brn{z-z_0}\fvv^+\brn{\kvv_\text{t}} +
\text{H}\brn{z_0-z}\fvv^-\brn{\kvv_\text{t}}} +
B_1\brn{\kvv_\text{t}, z} \gvv\brn{\kvv_\text{t}}\\[5mm]
\eta\Hvv^{(\text{d})}\brn{\kvv_\text{t}, z} & =
\text{sgn}\brn{z-z_0}\nonumber \\
& \cdot\brh{ A_1\brn{\kvv_\text{t}, z} \gvv\brn{\kvv_\text{t}}-
B_1\brn{\kvv_\text{t},
z}\brh{\text{H}\brn{z-z_0}\fvv^+\brn{\kvv_\text{t}} +
\text{H}\brn{z_0-z}\fvv^-\brn{\kvv_\text{t}}} }
\end{align}
where $\text{H}\brn{}$ denotes the Heaviside step-function,
$\text{sgn}\brn{}$ denotes the signum-function, and where for the
principal directions of the dipole
\begin{align}
\pvv=p\ux \Rightarrow  & \begin{cases}A_1\brn{\kvv_\text{t},z} = \ds
- \frac{\j p k_x}{2 \ep} \e^{\j\vs{k}_\text{t}\cdot\vs{\rho}_0} \e^{-\j k_z\brb{z-z_0}} \\[3mm]
B_1\brn{\kvv_\text{t},z}  =  \ds\frac{\j p k k_y}{2 \ep k_z}
\e^{\j\vs{k}_\text{t}\cdot\vs{\rho}_0} \e^{-\j k_z\brb{z-z_0}}
\end{cases} \label{eq:AB1x} \\[2mm]
\pvv=p\uy \Rightarrow  & \begin{cases}A_1\brn{\kvv_\text{t},z}  =
\ds - \frac{\j p k_y}{2 \ep}
\e^{\j\vs{k}_\text{t}\cdot\vs{\rho}_0} \e^{-\j k_z\brb{z-z_0}} \\[3mm]
B_1\brn{\kvv_\text{t},z}  =  \ds -\frac{\j p k k_x}{2 \ep k_z}
\e^{\j\vs{k}_\text{t}\cdot\vs{\rho}_0} \e^{-\j k_z\brb{z-z_0}}
\end{cases} \label{eq:AB1y} \\[2mm]
\pvv=p\uz \Rightarrow  & \begin{cases}A_1\brn{\kvv_\text{t},z}  =
\ds \text{sgn}\brn{z-z_0} \frac{\j p k_\text{t}^2}{2 \ep k_z}
\e^{\j\vs{k}_\text{t}\cdot\vs{\rho}_0} \e^{-\j k_z\brb{z-z_0}} \\[3mm]
B_1\brn{\kvv_\text{t},z}  =  0
\end{cases} \label{eq:AB1z}
\end{align}
When enforcing the boundary conditions in a multilayered structure,
it suffices to consider the transversal fields. For the dipole, the
transversal fields become
\begin{align}
\Evv_\text{t}^{(\text{d})}\brn{\kvv_\text{t}, z} & =
A_1\brn{\kvv_\text{t}, z}\fvv\brn{\kvv_\text{t}}  +
B_1\brn{\kvv_\text{t}, z} \gvv\brn{\kvv_\text{t}}\label{eq:Et:dip}\\[5mm]
\eta\Hvv_\text{t}^{(\text{d})}\brn{\kvv_\text{t}, z} & =
\text{sgn}\brn{z-z_0} \brh{ A_1\brn{\kvv_\text{t}, z}
\gvv\brn{\kvv_\text{t}}- B_1\brn{\kvv_\text{t},
z}\fvv\brn{\kvv_\text{t}} } \label{eq:Ht:dip}
\end{align}

The transversal fields originating from the charges and currents on
the ground plane at $z=0$ decay away from the ground plane and are for $z > 0$ thus written
\begin{align}
\Evv_\text{t}^{(\text{0})}\brn{\kvv_\text{t}, z} & =
\brh{A_0\brn{\kvv_\text{t}}\fvv\brn{\kvv_\text{t}}  +
B_0\brn{\kvv_\text{t}} \gvv\brn{\kvv_\text{t}}}\e^{-\j k_z z} \label{eq:Et:plane} \\[5mm]
\eta\Hvv_\text{t}^{(\text{0})}\brn{\kvv_\text{t}, z} & = \brh{
A_0\brn{\kvv_\text{t}} \gvv\brn{\kvv_\text{t}}-
B_0\brn{\kvv_\text{t}}\fvv\brn{\kvv_\text{t}} } \e^{-\j k_z z}
\label{eq:Ht:plane}
\end{align}

The spectral transversal fields originating from the charges and
currents on the strip in the plane $z=a$ are written
\begin{align}
\Evv_\text{t}^{(\text{s})}\brn{\kvv_\text{t}, z} & =
\brh{A_2\brn{\kvv_\text{t}}\fvv\brn{\kvv_\text{t}}  +
B_2\brn{\kvv_\text{t}} \gvv\brn{\kvv_\text{t}}}\e^{-\j k_z \brb{z-a}}\label{eq:Et:strip}\\[5mm]
\eta\Hvv_\text{t}^{(\text{s})}\brn{\kvv_\text{t}, z} & =
\text{sgn}\brn{z-a}\brh{ A_2\brn{\kvv_\text{t}}
\gvv\brn{\kvv_\text{t}}-
B_2\brn{\kvv_\text{t}}\fvv\brn{\kvv_\text{t}} } \e^{-\j k_z
\brb{z-a}} \label{eq:Ht:strip}
\end{align}
Note that since
$\fvv\brn{\kvv_\text{t}}\cdot\gvv\brn{\kvv_\text{t}}=0$,
\eqref{eq:Et:strip} is the unique Ansatz to enforce the continuity
of $\Evv_\text{t}^{(\text{s})}\brn{\kvv_\text{t}, z}$ across the
plane $z=a$.

\subsection{Boundary conditions}

The free surface current density $\Kvv\brn{\rhov}$ on the strip has
the corresponding spectral surface current density
\begin{align}
\Kvv\brn{\kvv_\text{t}} = \int_{\cal
S}\Kvv\brn{\rhov}\e^{\j\vs{k}_\text{t}\cdot\vs{\rho}}\dS
\label{eq:BC:Ht:1}
\end{align}
Applying the boundary condition
$\Hvv_\text{t}^{(\text{s})}\brn{\kvt,
z=a^+}-\Hvv_\text{t}^{(\text{s})}\brn{\kvt,
z=a^-}=\Kvv\brn{\kvt}\times\uz$, we obtain from \eqref{eq:Ht:strip}
that
\begin{align}
& 2 A_2\gvv - 2 B_2\fvv = \eta\Kvv\times\uz \label{eq:BC:Ht:2}
\end{align}
Utilizing \eqref{eq:modes:f}, \eqref{eq:modes:g} and
\eqref{eq:modes:fgort}, (\ref{eq:BC:Ht:2}) yields
\begin{align}
A_2 = \frac{\gvv\cdot\br{\eta\Kvv\times\uz}}{2 g^2} =
\frac{\br{\uz\times\gvv}\cdot\br{\eta\Kvv}}{2 g^2} =
-\frac{\kvv_\text{t}\cdot\br{\eta\Kvv}}{2k} = - \eta\frac{k_x K_x +
k_y K_y}{2 k} \label{eq:A2}
\\[3mm]
B_2 = -\frac{\fvv\cdot\br{\eta\Kvv\times\uz}}{2 f^2}
 = -\frac{\br{\uz\times\fvv}\cdot\br{\eta\Kvv}}{2 f^2} =
-\frac{\br{\uz\times\kvv_\text{t}}\cdot\br{\eta\Kvv}}{2 k_z} =-
\eta\frac{k_x K_y - k_y K_x}{2 k_z} \label{eq:B2}
\end{align}
The condition
$\Evv_\text{t}=\Evv_\text{t}^{(\text{d})}+\Evv_\text{t}^{(\text{0})}+\Evv_\text{t}^{(\text{s})}=\vc{0}$
at the plane $z=0$, together with \eqref{eq:modes:fgort},
\eqref{eq:Et:dip}, \eqref{eq:Et:plane} and \eqref{eq:Et:strip},
yields
\begin{align}
A_0 = - A_1\brn{z=0} - A_2\e^{-\j k_z a} \label{eq:A0}\\
B_0 = - B_1\brn{z=0} - B_2\e^{-\j k_z a} \label{eq:B0}
\end{align}

Thus, using \eqref{eq:Et:dip}, \eqref{eq:Et:plane},
\eqref{eq:Et:strip}, \eqref{eq:A2}-\eqref{eq:B0}, \eqref{eq:modes:f}
and \eqref{eq:modes:g}, we obtain that in the plane $z=a$ the
transversal spectral electric field becomes
\begin{align}
\Evv_{\text{t}a}\brn{\kvv_\text{t}} & = \brh{A_1\brn{\kvv_\text{t},
a} - A_1\brn{\kvv_\text{t}, 0}\e^{-\j k_z a} +
A_2\brn{\kvv_\text{t}}\br{1-\e^{-\j 2 k_z a
}}}\fvv\brn{\kvv_\text{t}}\nonumber \\
&  + \brh{B_1\brn{\kvv_\text{t}, a} - B_1\brn{\kvv_\text{t},
0}\e^{-\j k_z a} + B_2\brn{\kvv_\text{t}}\br{1-\e^{-\j 2 k_z a
}}}\gvv\brn{\kvv_\text{t}}
\nonumber \\
& = \Big\{\eta\br{1-\e^{-\j 2 k_z a}}\frac{k_x k_y
K_y\brn{\kvv_\text{t}} - \br{k^2-k_x^2}K_x\brn{\kvv_\text{t}}}{2 k
k_z} \nonumber \\& \quad + \frac{k_x k_z\br{A_1\brn{\kvv_\text{t},a}
- A_1\brn{\kvv_\text{t},0}\e^{-\j k_z a}} + k
k_y\br{-B_1\brn{\kvv_\text{t},a} +
B_1\brn{\kvv_\text{t},0}\e^{-\j k_z a} }}{k_x^2+k_y^2} \Big\}\ux \nonumber \\
& + \Big\{\eta\br{1-\e^{-\j 2 k_z a}}\frac{k_x k_y
K_x\brn{\kvv_\text{t}} - \br{k^2-k_y^2}K_y\brn{\kvv_\text{t}}}{2 k
k_z} \nonumber \\& \quad + \e^{-\j k_z a}\frac{k_y
k_z\br{A_1\brn{\kvv_\text{t},a} - A_1\brn{\kvv_\text{t},0}\e^{-\j
k_z a}} + k k_x\br{B_1\brn{\kvv_\text{t},a} -
B_1\brn{\kvv_\text{t},0}\e^{-\j k_z a}}}{k_x^2+k_y^2} \Big\}\uy
\label{eq:strip:Etkxky}
\end{align}

\subsection{Spectral surface current densities}

In the plane $z=a$, the spatial surface current density
$\Kvv\brn{x,y}$ is confined to the strip-conductor, with edges at $y
= \pm h$. Explicitly taking into account the singular behaviour at
the edges (see e.g. formula (4.13) in \cite{VanBladel}), the
components of the surface current density are expanded as
\begin{align}
K_x\brn{x,y} & =
\frac{\text{H}\brn{h-\brb{y}}}{\sqrt{1-y^2/h^2}}\sum_{n=0}^\infty
c_n\brn{x}\text{T}_n\brn{\ds\frac{y}{h}} \label{eq:Kx:spat}\\
K_y\brn{x,y} & =
\text{H}\brn{h-\brb{y}}\sqrt{1-y^2/h^2}\sum_{n=0}^\infty
d_n\brn{x}\text{U}_n\brn{\ds\frac{y}{h}} \nonumber \\
& =
\frac{\text{H}\brn{h-\brb{y}}}{2\sqrt{1-y^2/h^2}}\sum_{n=0}^\infty
d_n\brn{x}\brh{ \text{T}_n\brn{\ds\frac{y}{h}} -
\text{T}_{n+2}\brn{\ds\frac{y}{h}} } \label{eq:Ky:spat}
\end{align}
where $\text{T}_n\brn{}$ and $\text{U}_n\brn{}$ denote the $n$:th
order Chebyshev polynomials of first and second kind, respectively.
$\brc{c_n\brn{x}, d_n\brn{x}}_{n=0}^\infty$ are coefficient
functions to be determined.

Inserting (\ref{eq:Kx:spat}) and (\ref{eq:Ky:spat}) into
(\ref{eq:BC:Ht:1}) and using formula 7.355 in
\cite{GradshteynRyzhik}, the components of the spectral surface
current density become
\begin{align}
K_x\brn{\kvv_\text{t}} &  = \sum_{n=0}^\infty
c_n\brn{k_x}\int_{-h}^h
\frac{\text{T}_n\brn{y/h}}{\sqrt{1-y^2/h^2}}\e^{\j k_y y}\dd y \nonumber \\
& = \pi h \sum_{n=0}^\infty \br{-1}^n
\brh{c_{2n}\brn{k_x}\bJ{2n}{k_y h} + \j\:
c_{2n+1}\brn{k_x}\bJ{2n+1}{k_y
h} } \label{eq:Kx:spectral}\\[5mm]
K_y\brn{\kvv_\text{t}} &  = \frac{1}{2}\sum_{n=0}^\infty
d_n\brn{k_x}\int_{-h}^h
\frac{\text{T}_n\brn{y/h}-\text{T}_{n+2}\brn{y/h}}{\sqrt{1-y^2/h^2}}\e^{\j
k_y y}\dd y = \nonumber \\
& = \frac{\pi h}{2} \sum_{n=0}^\infty \br{-1}^n \Big[
d_{2n}\brn{k_x}\brc{\bJ{2n}{k_y h} +  \bJ{2n+2}{k_y h}} \nonumber
\\
& \qquad\qquad\qquad + \j\: d_{2n+1}\brn{k_x}\brc{ \bJ{2n+1}{k_y h}
+ \bJ{2n+3}{k_y h} } \Big] \nonumber \\
& = \pi h \sum_{n=0}^\infty \br{-1}^n \brh{
d_{2n}\brn{k_x}\frac{\br{2n+1}\bJ{2n+1}{k_y h}}{k_y h} + \j\:
d_{2n+1}\brn{k_x}\frac{\br{2n+2}\bJ{2n+2}{k_y h}}{k_y h} }
\label{eq:Ky:spectral}
\end{align}
where the spectral coefficients
\begin{align}
c_n\brn{k_x} = \int_{-\infty}^\infty c_n\brn{x} \e^{\j k_x x} \dd x,
\qquad d_n\brn{k_x} = \int_{-\infty}^\infty d_n\brn{x} \e^{\j k_x x}
\dd x   \label{eq:Kxy:FTcoeff}
\end{align}
and $\text{J}_n\brn{}$ denotes the $n$:th order Bessel function.

\subsection{Equation system for the spectral coefficients}

In the plane $z=a$, the spatial transversal electric field becomes
\begin{align}
\Evv_{\text{t}a}\brn{x,y} = \frac{1}{4\pi^2} \int_{-\infty}^\infty
\e^{-\j k_x^\prime x} \dd k_x^\prime  \int_{-\infty}^\infty \e^{-\j
k_y y} \dd k_y \Evv_{\text{t}a}\brn{k_x^\prime,
k_y}\label{eq:strip:Etxy}
\end{align}


On the strip, $\br{-\infty < x < \infty}\cap \br{ -h < y < h}\cap
\br{z=a}$, we have $\Evv_{\text{t}a}\brn{x,y}=\vc{0}$. According to
equation (4.9) in \cite{VanBladel}, it holds in the vicinity of an
edge that $E_x\propto \sqrt{r}, E_y\propto 1/\sqrt{r}$, where $r$ is
the distance to the edge. Taking into account these properties, we
enforce the vanishing tangential field on the strip through the
following testing procedures:
\begin{align}
\int_{-\infty}^\infty \e^{\j k_x x} \dd x \int_{-h}^h
\frac{\text{T}_m\brn{y/h}}{\sqrt{1-y^2/h^2}} E_x\brn{x,y} \dd y = 0,
\quad -\infty < k_x < \infty, \quad m=0, 1, \ldots
\label{eq:strip:testEx}
\\
\int_{-\infty}^\infty \e^{\j k_x x} \dd x \int_{-h}^h
\sqrt{1-y^2/h^2}\text{U}_m\brn{y/h} E_y\brn{x,y} \dd y = 0, \quad
-\infty < k_x < \infty, \quad m=0, 1, \ldots \label{eq:strip:testEy}
\end{align}
When inserting (\ref{eq:strip:Etxy}) into (\ref{eq:strip:testEx})
and (\ref{eq:strip:testEy}), the spatial integrals result in
\begin{align}
& \frac{1}{2\pi}\int_{-\infty}^\infty \e^{\j\br{k_x - k_x^\prime}x}
\dd x = \dirac{k_x - k_x^\prime} \\[2mm]
& \int_{-h}^h \frac{\text{T}_m\brn{y/h}}{\sqrt{1-y^2/h^2}} \e^{-\j
k_y y}\dd y \propto \bJ{m}{k_y h} \\[2mm]
& \int_{-h}^h \sqrt{1-y^2/h^2}\text{U}_m\brn{y/h} \e^{-\j k_y y}\dd
y \propto \br{m+1}\frac{\bJ{m+1}{k_y h}}{k_y h}
\end{align}
Hence, (\ref{eq:strip:testEx}) and (\ref{eq:strip:testEy}) imply the
following results in the spectral domain:
\begin{align}
& \int_{-\infty}^\infty E_x\brn{k_x, k_y} \bJ{m}{k_y h} \dd k_y = 0,
\quad m=0, 1, \ldots \label{eq:strip:Ex=0:spectral}
\\[2mm]
& \int_{-\infty}^\infty E_y\brn{k_x, k_y} \frac{\bJ{m+1}{k_y h}}{k_y
h} \dd k_y = 0, \quad m=0, 1, \ldots
\end{align}
which, using \eqref{eq:strip:Etkxky} and
\eqref{eq:AB1x}-\eqref{eq:AB1z}, become
\begin{align}
\int_{-\infty}^\infty & \br{1-\e^{-\j 2 k_z a}}\frac{
\br{k^2-k_x^2}K_x\brn{\kvv_\text{t}} - k_x k_y
K_y\brn{\kvv_\text{t}} }{2 k k_z} \bJ{m}{k_y h} \dd k_y\nonumber
\\ 
& = \frac{\j p }{2\ep\eta}\e^{\j k_x x_0} \int_{-\infty}^\infty
\e^{\j k_y y_0} \bJ{m}{k_y h} \dd k_y \left\{\begin{array}{rr}
-\ds\frac{k^2-k_x^2}{k_z}\brh{\e^{-\j k_z\brb{a-z_0}} - \e^{-\j
k_z\br{a+z_0}}}  &: \pvv=p\ux \\[5mm]
\ds\frac{k_x k_y}{k_z}\brh{\e^{-\j k_z\brb{a-z_0}} - \e^{-\j
k_z\br{a+z_0}}}  &: \pvv=p\uy \\[5mm]
k_x\brh{\text{sgn}\brn{a-z_0}\e^{-\j k_z\brb{a-z_0}} + \e^{-\j
k_z\br{a+z_0}}}  &: \pvv=p\uz
\end{array}\right. \label{eq:strip:eq1}
\end{align}
\begin{align}
\int_{-\infty}^\infty & \br{1-\e^{-\j 2 k_z a}}\frac{
\br{k^2-k_y^2}K_y\brn{\kvv_\text{t}} -
k_x k_y K_x\brn{\kvv_\text{t}} }{2 k k_z} \frac{\bJ{m+1}{k_y h}}{k_y h}\dd k_y\nonumber \\
& = \frac{\j p }{2\ep\eta}\e^{\j k_x x_0} \int_{-\infty}^\infty
\e^{\j k_y y_0} \bJ{m+1}{k_y h} \dd k_y \left\{\begin{array}{rr}
\ds\frac{k_x}{k_z h}\brh{\e^{-\j k_z\brb{a-z_0}} - \e^{-\j
k_z\br{a+z_0}}}  &: \pvv=p\ux \\[5mm]
-\ds\frac{k^2-k_y^2}{k_y k_z h}\brh{\e^{-\j k_z\brb{a-z_0}} -
\e^{-\j
k_z\br{a+z_0}}}  &: \pvv=p\uy \\[5mm]
\ds\frac{1}{h}\brh{\text{sgn}\brb{a-z_0}\e^{-\j k_z\brb{a-z_0}} +
\e^{-\j k_z\br{a+z_0}}} &: \pvv=p\uz
\end{array}\right. \label{eq:strip:eq2}
\end{align}
Next, we insert \eqref{eq:Kx:spectral} and \eqref{eq:Ky:spectral}
into (\ref{eq:strip:eq1}) and (\ref{eq:strip:eq2}), and utilize the
even and odd parities of integer order Bessel functions, to obtain
the following systems for $\brc{c_{2n}\brn{k_x},
d_{2n+1}\brn{k_x}}_{n=0}^\infty$:
\begin{align}
& \pi h \sum_{n=0}^\infty \br{-1}^n\int_0^\infty \frac{1-\e^{-\j 2
k_z a}}{k k_z} \bigg\{ \br{k^2-k_x^2}\bJ{2m}{k_y h} \bJ{2n}{k_y h}
c_{2n}\brn{k_x}
\nonumber \\
& \text{\hspace{60mm}} - \j\frac{k_x}{h} \br{2n+2} \bJ{2m}{k_y h}
\bJ{2n+2}{k_y h} d_{2n+1}\brn{k_x} \bigg\} \dd k_y \nonumber \\
& = \frac{\j p }{\ep\eta}\e^{\j k_x x_0} \int_0^\infty  \bJ{2m}{k_y
h} \dd k_y \left\{\begin{array}{rr}
-\ds\frac{k^2-k_x^2}{k_z}\brh{\e^{-\j k_z\brb{a-z_0}} - \e^{-\j
k_z\br{a+z_0}}} \cos\brn{k_y y_0}  &: \pvv=p\ux \\[5mm]
\ds\frac{k_x k_y}{k_z}\brh{\e^{-\j k_z\brb{a-z_0}} - \e^{-\j
k_z\br{a+z_0}}}\j\sin\brn{k_y y_0}  &: \pvv=p\uy \\[5mm]
k_x\brh{\text{sgn}\brn{a-z_0}\e^{-\j k_z\brb{a-z_0}} + \e^{-\j
k_z\br{a+z_0}}}\cos\brn{k_y y_0}  &: \pvv=p\uz
\end{array}\right.  \label{eq:strip:cjdu1}
\end{align}
\begin{align}
& \pi h \sum_{n=0}^\infty \br{-1}^n\int_0^\infty \frac{1-\e^{-\j 2
k_z a}}{k k_z} \bigg\{ \j\frac{k^2-k_y^2}{k_y^2
h^2}\br{2n+2}\bJ{2m+2}{k_y h} \bJ{2n+2}{k_y h} d_{2n+1}\brn{k_x}
\nonumber \\
& \text{\hspace{60mm}} -\frac{k_x}{h}\bJ{2m+2}{k_y h}
\bJ{2n}{k_y h} c_{2n}\brn{k_x} \bigg\} \dd k_y \nonumber \\
& = \frac{\j p }{\ep\eta}\e^{\j k_x x_0} \int_0^\infty \bJ{2m+2}{k_y
h} \dd k_y \left\{\begin{array}{rr} \ds\frac{k_x}{k_z h}\brh{\e^{-\j
k_z\brb{a-z_0}} - \e^{-\j
k_z\br{a+z_0}}}\cos\brn{k_y y_0}  &: \pvv=p\ux \\[5mm]
-\ds\frac{k^2-k_y^2}{k_y k_z h}\brh{\e^{-\j k_z\brb{a-z_0}} -
\e^{-\j
k_z\br{a+z_0}}}\j\sin\brn{k_y y_0}  &: \pvv=p\uy \\[5mm]
\ds\frac{1}{h}\brh{\text{sgn}\brn{a-z_0}\e^{-\j k_z\brb{a-z_0}} +
\e^{-\j k_z\br{a+z_0}}}\cos\brn{k_y y_0} &: \pvv=p\uz
\end{array}\right. \label{eq:strip:cjdu2}
\end{align}
and into the following systems for $\brc{c_{2n+1}\brn{k_x},
d_{2n}\brn{k_x}}_{n=0}^\infty$:
\begin{align}
& \pi h \sum_{n=0}^\infty \br{-1}^n\int_0^\infty \frac{1-\e^{-\j 2
k_z a}}{k k_z} \bigg\{ \j\br{k^2-k_x^2}\bJ{2m+1}{k_y h}
\bJ{2n+1}{k_y h} c_{2n+1}\brn{k_x}
\nonumber \\
& \text{\hspace{60mm}} - \frac{k_x}{h} \br{2n+1} \bJ{2m+1}{k_y h}
\bJ{2n+1}{k_y h} d_{2n}\brn{k_x} \bigg\} \dd k_y \nonumber \\
& = \frac{\j p }{\ep\eta}\e^{\j k_x x_0} \int_0^\infty \bJ{2m+1}{k_y
h} \dd k_y \left\{\begin{array}{rr}
-\ds\frac{k^2-k_x^2}{k_z}\brh{\e^{-\j k_z\brb{a-z_0}} - \e^{-\j
k_z\br{a+z_0}}} \j\sin\brn{k_y y_0}  &: \pvv=p\ux \\[5mm]
\ds\frac{k_x k_y}{k_z}\brh{\e^{-\j k_z\brb{a-z_0}} - \e^{-\j
k_z\br{a+z_0}}}\cos\brn{k_y y_0}  &: \pvv=p\uy \\[5mm]
k_x\brh{\text{sgn}\brn{a-z_0}\e^{-\j k_z\brb{a-z_0}} + \e^{-\j
k_z\br{a+z_0}}}\j\sin\brn{k_y y_0}  &: \pvv=p\uz
\end{array}\right. \label{eq:strip:cudj1}
\end{align}
\begin{align}
& \pi h \sum_{n=0}^\infty \br{-1}^n\int_0^\infty \frac{1-\e^{-\j 2
k_z a}}{k k_z} \bigg\{ \frac{k^2-k_y^2}{k_y^2
h^2}\br{2n+1}\bJ{2m+1}{k_y h} \bJ{2n+1}{k_y h} d_{2n}\brn{k_x}
\nonumber \\
& \text{\hspace{60mm}} -\j\frac{k_x}{h}\bJ{2m+1}{k_y h}
\bJ{2n+1}{k_y h} c_{2n+1}\brn{k_x} \bigg\} \dd k_y \nonumber \\
& = \frac{\j p }{\ep\eta}\e^{\j k_x x_0} \int_0^\infty \bJ{2m+1}{k_y
h} \dd k_y \left\{\begin{array}{rr} \ds\frac{k_x}{k_z h}\brh{\e^{-\j
k_z\brb{a-z_0}} - \e^{-\j
k_z\br{a+z_0}}}\j\sin\brn{k_y y_0}  &: \pvv=p\ux \\[5mm]
-\ds\frac{k^2-k_y^2}{k_y k_z h}\brh{\e^{-\j k_z\brb{a-z_0}} -
\e^{-\j
k_z\br{a+z_0}}}\cos\brn{k_y y_0}  &: \pvv=p\uy \\[5mm]
\ds\frac{1}{h}\brh{\text{sgn}\brn{a-z_0}\e^{-\j k_z\brb{a-z_0}} +
\e^{-\j k_z\br{a+z_0}}}\j\sin\brn{k_y y_0} &: \pvv=p\uz
\end{array}\right.  \label{eq:strip:cudj2}
\end{align}
Summarizing, (\ref{eq:strip:cjdu1})-(\ref{eq:strip:cudj2}) are
infinitely large linear equation systems for the spectral
coefficients $c_n\brn{k_x}, d_n\brn{k_x}$. The elements in the
system matrices and driving terms are obtained through various
integrals with respect to the spectral variable $k_y$, with the
other spectral variable $k_x$ as a parameter. 

For the numerical solution,
(\ref{eq:strip:cjdu1})-(\ref{eq:strip:cudj2}) are truncated into a
finite number of coefficients and testing functions, and solved for
different values of $k_x$. With the spectral coefficients
determined, the spectral surface current density follows from
(\ref{eq:Kx:spectral}) and (\ref{eq:Ky:spectral}), the mode
coefficients from \eqref{eq:A2}-\eqref{eq:B0}, and the spectral
fields from (\ref{eq:modes:TM_E})-(\ref{eq:modes:TE_H}). The details
of the numerical evaluation of the spectral integrals are described
in Appendix \ref{section:app:integrals}. Finally, the spatial fields
follow from (\ref{eq:modes:E_spatial}) and
(\ref{eq:modes:H_spatial}).

The coefficients for the spatial surface current density follow from
the inverse Fourier transform of (\ref{eq:Kxy:FTcoeff}):
\begin{align}
c_n\brn{x} = \frac{1}{2\pi}\int_{-\infty}^\infty c_n\brn{k_x}
\e^{-\j k_x x} \dd k_x, \qquad d_n\brn{x} =
\frac{1}{2\pi}\int_{-\infty}^\infty d_n\brn{k_x} \e^{-\j k_x x} \dd
k_x \label{eq:strip:cd:spat}
\end{align}
With $\text{T}_0 = 1$, it follows from (\ref{eq:Kx:spat}) and the
orthogonality relation for the Chebyshev polynomials that the total
current flowing along the strip becomes
\begin{align}
I\brn{x} & = \int_{-\infty}^\infty K_x\brn{x,y} \dd y  = \int_{-h}^h
\frac{\dd y}{\sqrt{1-y^2/h^2}}\sum_{n=0}^\infty
c_n\brn{x}\text{T}_n\brn{\ds\frac{y}{h}} \nonumber \\
& = h \sum_{n=0}^\infty c_n\brn{x} \int_{-1}^1
\frac{\text{T}_0\brn{u}\text{T}_n\brn{u}}{\sqrt{1-u^2}}  \dd u= h
\pi \sum_{n=0}^\infty c_n\brn{x} \updelta_{n0} = \pi h
c_0\brn{x}\label{eq:strip:Ix}
\end{align}

\section{Numerical results}

\label{section:Numerical_results}

In this section, we evaluate the method through several numerical
examples. In all examples, the dipole source has the $x$-coordinate
$x_0=0$ m, the frequency $f=300$ MHz, and the medium parameters
$\ep=\ep_0\br{1-\j 10^{-5}}, \mu=\mu_0\br{1-\j 10^{-5}}$. Hence, it
follows that the wavelength $\lambda\approx 1$ m and that
attenuation due to medium losses have minor impact within distances
of the order of some hundred meters\footnote{In more realistic cases, attenuation is due to a finite ground conductivity that will mask the losses in the air. The simple PEC-model serves to not obscure the main results with too massive algebra.}.

\subsection{Surface currents on a wide strip }

Here, we consider a comparatively wide strip with the half-width
$h=0.5$ m. The strip is at the height $a=1$ m and the dipole source
is at the height $z_0 = 0.5$ m, above the PEC plane. The required
number of spectral coefficients scales with $h/\lambda$ and here we
use $m_\text{max}=3$ in
(\ref{eq:strip:cjdu1})-(\ref{eq:strip:cudj2}), i.e. the sought
coefficients are $\brc{c_n, d_n}_{n=0}^7$.

For graphical clarity and space reasons, we only consider the
portion -3 m $< x <$ 3 m of the strip, and the results shown are
only for the real part (one particular instantaneous value) of the
surface current density.

In cases with the dipole moment in the $x$-direction, the field will
be purely TM in the {\em $x$-direction}, whereby $K_y = 0$. This
extra numerical check resulted in $K_y$-values that were about five
orders in magnitude smaller than the $K_x$-values. Hence, for
$x$-directed dipoles only results for $K_x$ are presented.

\begin{figure}[t]
\centering \subfigure[Dipole moment $\pvv=1\ux$ Cm. $K_y=0$.]{
\psfrag{x}{$x$/m} \psfrag{y}{$y$/m}
\psfrag{R}{$\ds\frac{\text{Re}\brc{K_x}}{\text{A/m}}$}
\label{Fig:num:wide:Bred1a}
\includegraphics[scale=0.9]{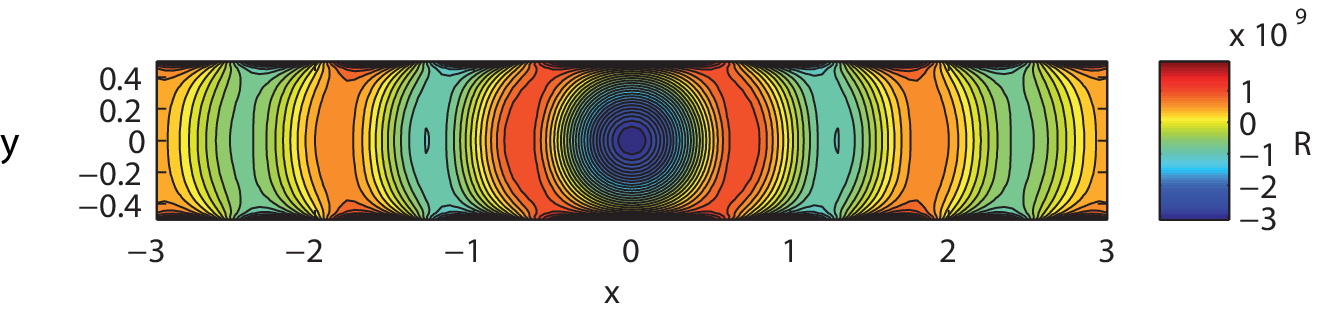} }
 \subfigure[Dipole moment $\pvv=1\uy$ Cm.]{ \psfrag{x}{$x$/m}
\psfrag{y}{$y$/m}
\psfrag{R}{$\ds\frac{\text{Re}\brc{K_x}}{\text{A/m}}$}
\psfrag{S}{$\ds\frac{\text{Re}\brc{K_y}}{\text{A/m}}$}
\label{Fig:num:wide:Bred1b}
\includegraphics[scale=0.9]{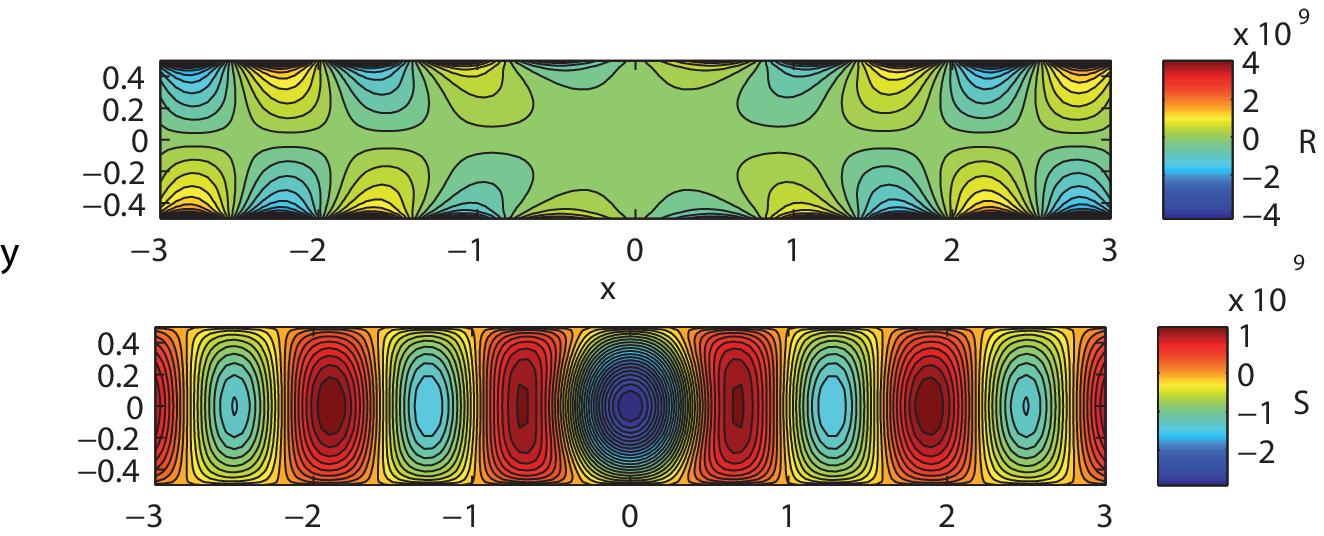} }
\subfigure[Dipole moment $\pvv=1\uz$ Cm.]{ \psfrag{x}{$x$/m}
\psfrag{y}{$y$/m}
\psfrag{R}{$\ds\frac{\text{Re}\brc{K_x}}{\text{A/m}}$}
\psfrag{S}{$\ds\frac{\text{Re}\brc{K_y}}{\text{A/m}}$}
\label{Fig:num:wide:Bred1c}
\includegraphics[scale=0.9]{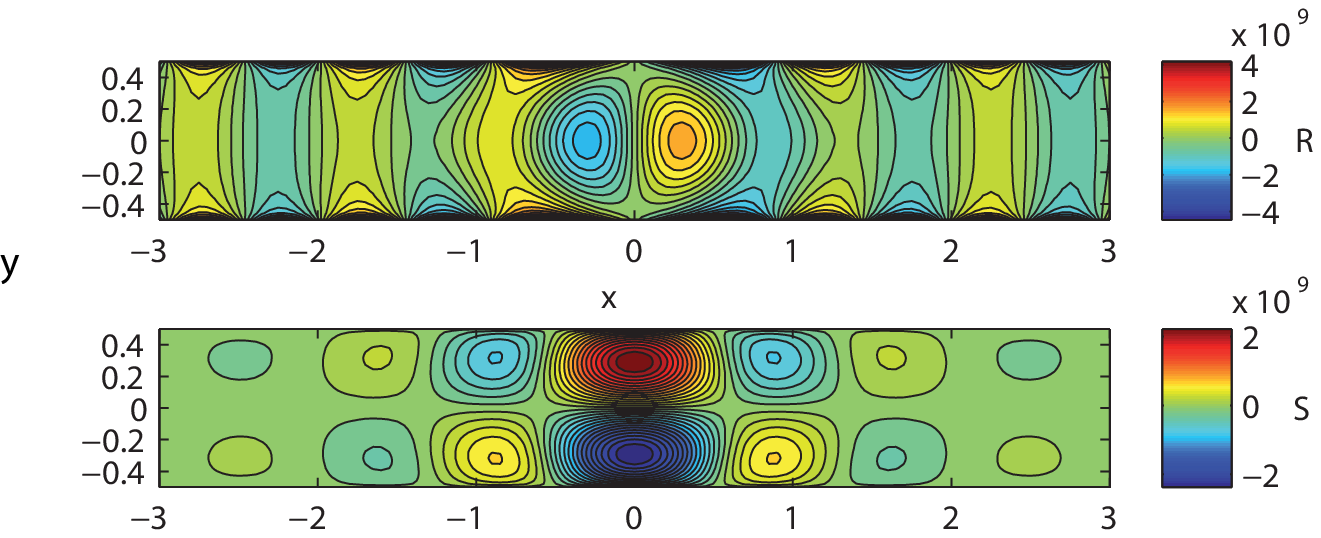} }
\caption{Surface current when $h=0.5$ m, $a=1$ m. Dipole at $\rvv_0
= 0.5\uz$ m.} \label{Fig:num:wide:Bred1}
\end{figure}

We check the symmetries expected when the dipole
is centered with the strip at $y_0=0$. 
The results are shown in
Figure \ref{Fig:num:wide:Bred1}.
We see that $K_x$ and $K_y$ exhibit the symmetries expected, due to
the locations and orientations of the dipole sources. In Figure
\ref{Fig:num:wide:Bred1b}, we see that the $y$-directed dipole
yields odd symmetry in $K_x$, resulting in a zero net current
flowing along the strip (cf. \eqref{eq:strip:Ix}). Also, $K_y$
dominates over $K_x$, since at the strip the exciting dipole
electric field is predominantly in the $y$-direction. For the
$z$-directed dipole, we see in Figure \ref{Fig:num:wide:Bred1c}
that, except for in the region above the dipole, $K_x$ dominates
over $K_y$.


\subsection{Surface currents on a narrow strip}

\label{subsection:K_on_narrow_strip}

Here, we consider a comparatively narrow strip with the half-width
$h=0.1$ m. The strip is at the height $a=6$ m and the dipole source
is at the height $z_0 = 5.5$ m, above the PEC plane, and
off-centered at $y_0 = 0.5$ m. Here we use $m_\text{max}=1$ in
(\ref{eq:strip:cjdu1})-(\ref{eq:strip:cudj2}), i.e. the sought
coefficients are $\brc{c_n, d_n}_{n=0}^3$.

\begin{figure}[t]
\centering \subfigure[Dipole moment $\pvv=1\uy$ Cm.]{
\psfrag{x}{$x$/m} \psfrag{y}{$y$/m}
\psfrag{R}{$\ds\frac{\text{Re}\brc{K_x}}{\text{A/m}}$}
\psfrag{S}{$\ds\frac{\text{Re}\brc{K_y}}{\text{A/m}}$}
\label{Fig:num:narrow:Smaltest3a}
\includegraphics[scale=0.9]{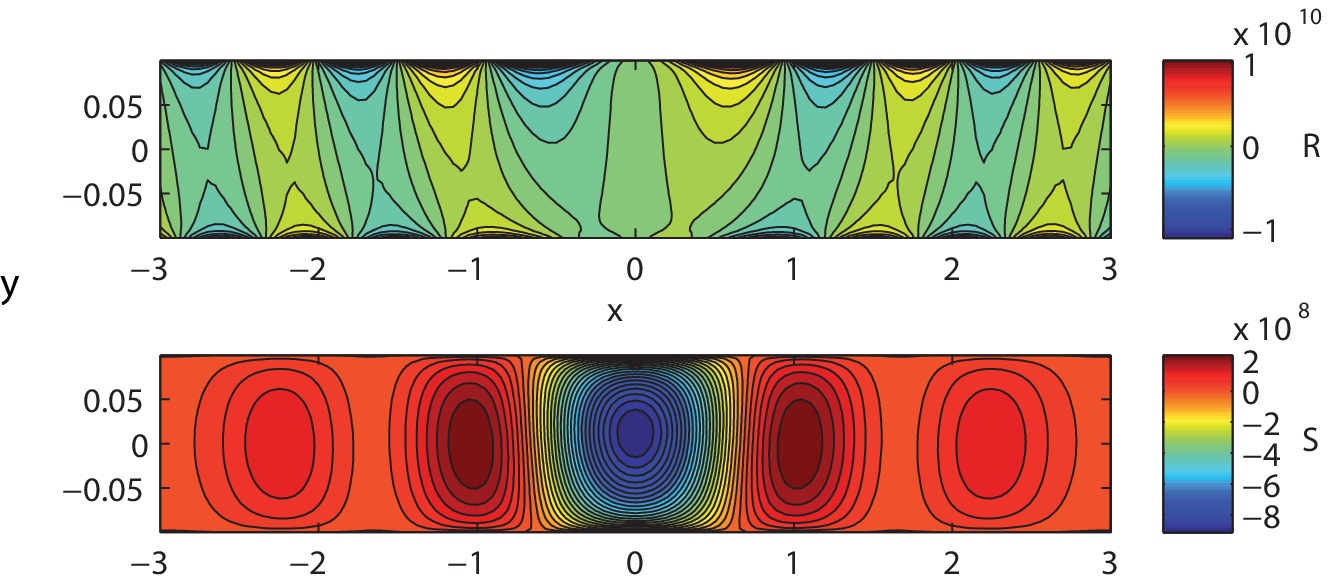} }
\subfigure[Dipole moment $\pvv=1\uz$ Cm.]{ \psfrag{x}{$x$/m}
\psfrag{y}{$y$/m}
\psfrag{R}{$\ds\frac{\text{Re}\brc{K_x}}{\text{A/m}}$}
\psfrag{S}{$\ds\frac{\text{Re}\brc{K_y}}{\text{A/m}}$}
\label{Fig:num:narrow:Smaltest3b}
\includegraphics[scale=0.9]{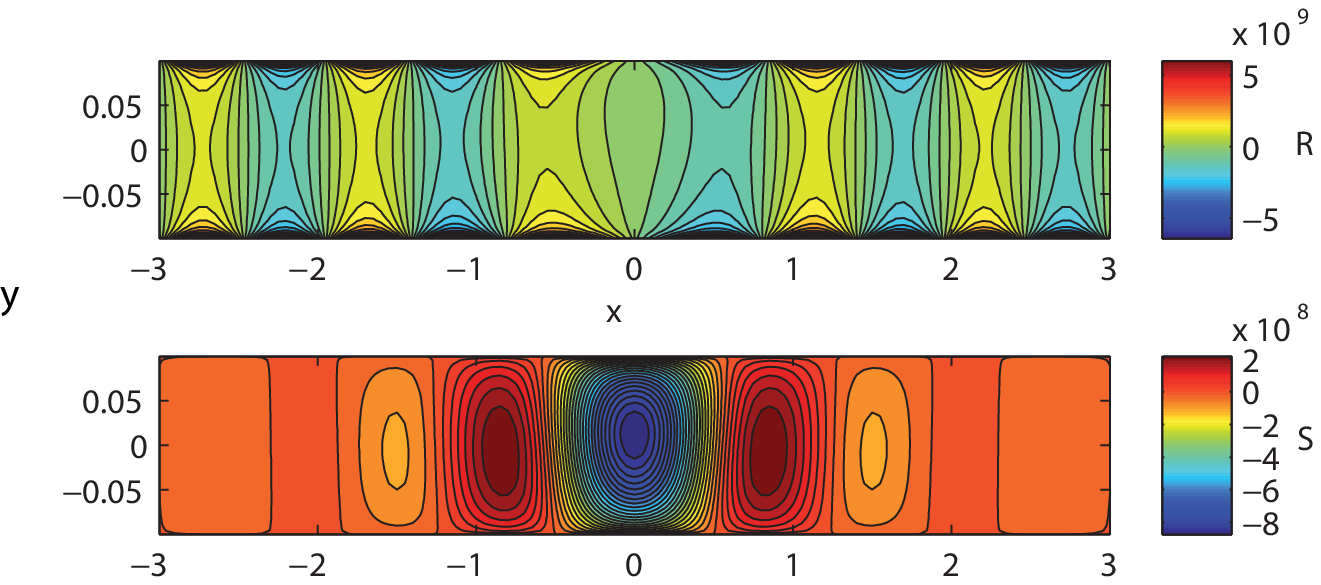} }
\caption{Surface current when $h=0.1$ m, $a=6$ m. Dipole at $\rvv_0
= 0.5\uy + 5.5\uz$ m. Note that, for graphical clarity, different
scalings are used on the $x$- and $y$-axes.}
\label{Fig:num:narrow:Smaltest3}
\end{figure}

The conjecture is that for a narrow strip the
longitudinal\footnote{Note that \cte{longitudinal} here refers to
the direction of the strip, the $x$-direction, while in Section
\ref{section:Problem_formulation_and_theory} \cte{longitudinal}
refers to the decomposition direction of the fields, the
$z$-direction.} current dominates over the transverse current, i.e.
$\brb{K_x}\gg \brb{K_y}$. Since an $x$-directed dipole yields
$K_y=0$, results for that case are omitted, due to space reasons.
The results for $y$- and $z$-directed dipoles, shown in
\ref{Fig:num:narrow:Smaltest3}
Figures \ref{Fig:num:narrow:Smaltest3a} and
\ref{Fig:num:narrow:Smaltest3b} respectively, show that for both the
$y$-directed and the $z$-directed dipole, $K_x$ dominates over
$K_y$, which supports the conjecture that on a narrow strip the
surface current density is predominantly longitudinal.

\subsection{Simplified model for narrow strips using only the total long\-itudinal currents}

Guided by the results in subsection
\ref{subsection:K_on_narrow_strip}, we will here make
approximations that result in a considerably simplified model.

First, the transverse surface current is neglected completely, i.e.
we set $K_y = 0$ whereby $\brc{d_n=0}_{n=0}^\infty$ (cf.
\eqref{eq:Ky:spat}). Next, assuming that the strip is not too close
to the PEC plane and not too close to the dipole, we neglect
proximity effects and assume that in \eqref{eq:Kx:spat} the
longitudinal surface current density can be described by the zeroth
order coefficient only, i.e. $\brc{c_n=0}_{n=1}^\infty$. From
\eqref{eq:Kx:spat} and \eqref{eq:strip:Ix}, it thus follows that
\begin{align}
K_x\brn{x,y} = \frac{I\brn{x}}{\pi h}
\frac{\text{H}\brn{h-\brb{y}}}{\sqrt{1-y^2/h^2}} \label{eq:Kx0:spat}
\end{align}
and, introducing the spectral current $I\brn{k_x}=\pi h
c_0\brn{k_x}$, that \eqref{eq:Kx:spectral} reduces to
\begin{align}
K_x\brn{\kvt} = I\brn{k_x} \bJ{0}{k_y h}
\end{align}
In order to avoid an over-determined system, the final approximation
is that we omit using \eqref{eq:strip:testEy} to enforce $E_y=0$ on
the strip. Hence, we only use $m=0$ in \eqref{eq:strip:testEx} to
enforce that the longitudinal electric field component $E_x=0$ on
the strip, whereby \eqref{eq:strip:Ex=0:spectral} reduces to
\begin{align}
\int_{-\infty}^\infty E_x\brn{k_x, k_y} \bJ{0}{k_y h}\dd k_y = 0
\end{align}
Now, under these approximations, the general equations
\eqref{eq:strip:cjdu2}-\eqref{eq:strip:cudj2} fall out, leaving us
with \eqref{eq:strip:cjdu1}, which, as a scalar equation for the
spectral current, becomes
\begin{align}
& I\brn{k_x} \frac{k^2-k_x^2}{k}\int_0^\infty \frac{1-\e^{-\j 2 k_z
a}}{k_z} \text{J}_0^2\brn{k_y h} \dd k_y \nonumber \\
& = \frac{\j p }{\ep\eta}\e^{\j k_x x_0} \int_0^\infty  \bJ{0}{k_y
h} \dd k_y \left\{\begin{array}{rr}
-\ds\frac{k^2-k_x^2}{k_z}\brh{\e^{-\j k_z\brb{a-z_0}} - \e^{-\j
k_z\br{a+z_0}}} \cos\brn{k_y y_0}  &: \pvv=p\ux \\[5mm]
\ds\frac{k_x k_y}{k_z}\brh{\e^{-\j k_z\brb{a-z_0}} - \e^{-\j
k_z\br{a+z_0}}}\j\sin\brn{k_y y_0}  &: \pvv=p\uy \\[5mm]
k_x\brh{\text{sgn}\brn{a-z_0}\e^{-\j k_z\brb{a-z_0}} + \e^{-\j
k_z\br{a+z_0}}}\cos\brn{k_y y_0}  &: \pvv=p\uz
\end{array}\right.  \label{eq:strip:c0}
\end{align}
To verify the approximation of using only the total longitudinal
current, we define the relative difference
$d\brn{x}=\brb{I_\text{app}\brn{x}-I_\text{gen}\brn{x}}/\brb{I_\text{gen}\brn{x}}$,
where $I_\text{app}\brn{x}$ is the current obtained from using the
approximate equation \eqref{eq:strip:c0} and $I_\text{gen}\brn{x}$
is the current obtained from using the complete system of equations.
In the comparison example, the geometrical parameters are the same
as in Figure \ref{Fig:num:narrow:Smaltest3}. With the dipole at
$x=0$, symmetry yields that we only need to consider $x\geq 0$, and
here we consider the range $0 \leq x \leq 40$ m.
\begin{figure}[t]
\centering \psfrag{x}{$\pvv=p\ux$} \psfrag{y}{$\pvv=p\uy$}
\psfrag{z}{$\pvv=p\uz$} \psfrag{m}{$x$/m}
\psfrag{d}{$d\brn{x}=\ds\frac{\brb{I_\text{app}\brn{x}-I_\text{gen}\brn{x}}}{\brb{I_\text{gen}\brn{x}}}$}
\includegraphics[scale=0.95]{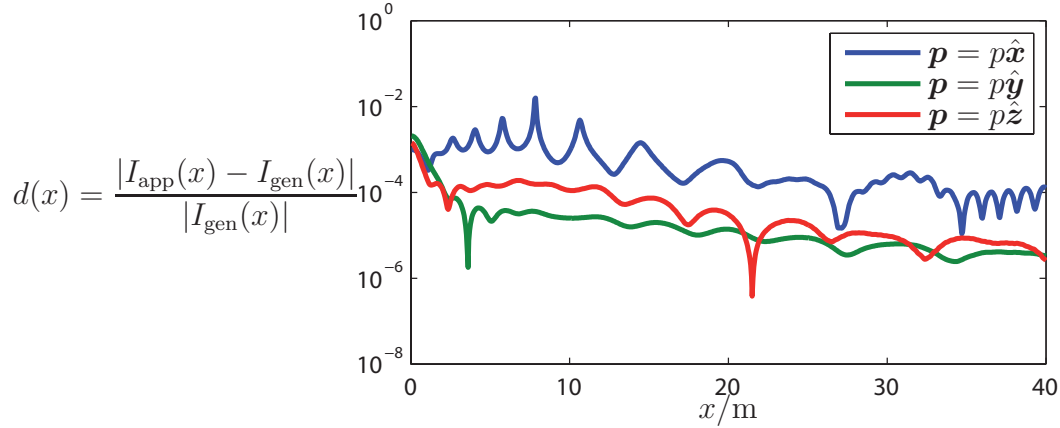}
\caption{Relative difference between currents obtained using the
approximate method and the general method.}
\label{Fig:num:narrow:IfullI0}
\end{figure}
The results are shown in Figure \ref{Fig:num:narrow:IfullI0}, where
we see that $d\brn{x}\ll 1$, even in this case with a strip as wide
as 1/5 of the wavelength.

\subsection{Results for narrow strips compared with results for thin circular wires}

In this section, we compare results for the narrow strip with results for the equivalent circular wire. With negligible proximity effects, the per length inductances and capacitances for a wire with radius $s$ become the same as for the strip with halfwidth $h$ if $s=h/2$; see e.g. \cite{Norgren2003}.

\subsubsection{Identification of TEM-mode in the $x$-direction}

\label{section:TEMcomp}

For a wire/strip at small height (in terms of wavelengths) above the ground plane, the TEM-mode (if excited) is expected to be the dominating mode, in the field and current distributions. Note that the propagation along the wire/strip is in the $x$-direction, wherefore TEM here refers to field components in the $y$- and $z$-directions only.

In Appendix \ref{section:app:TEMmod}, we derive the expressions \eqref{eq:TEM:Ix}-\eqref{eq:TEM:Iz} for the ideal TEM-mode currents on the wire excited by dipole sources in the principal directions.

\begin{figure}[t]
\centering
\subfigure[Dipole moment $\pvv=1\ux$ Cm.]{
\psfrag{x}{$x$/m} \psfrag{y}{$y$/m}
\psfrag{R}{$\ds\frac{\text{Re}\brc{K_x}}{\text{A/m}}$}
\psfrag{S}{$\ds\frac{\text{Re}\brc{K_y}}{\text{A/m}}$}
\label{Fig:num:ITEM:Pappfig2x}
\psfrag{I}{$\ds\frac{\text{Re}\brc{I}}{\text{A}}$} \psfrag{x}{$x$/m}
\includegraphics[scale=0.95]{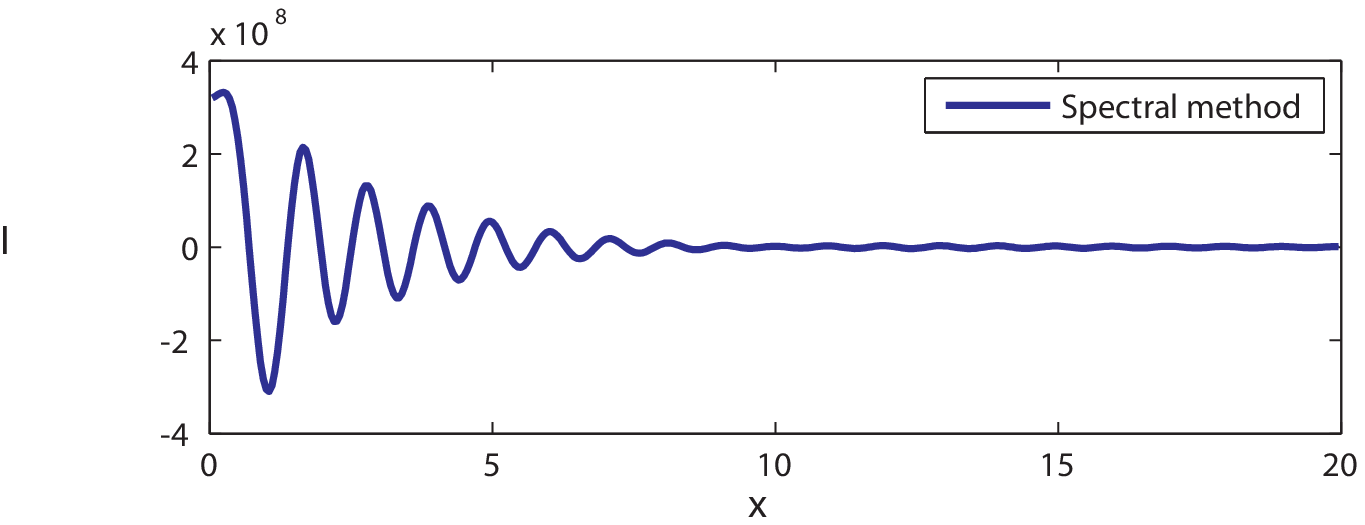} } 
\subfigure[Dipole moment $\pvv=1\uy$ Cm.]{
\psfrag{x}{$x$/m} \psfrag{y}{$y$/m}
\psfrag{R}{$\ds\frac{\text{Re}\brc{K_x}}{\text{A/m}}$}
\psfrag{S}{$\ds\frac{\text{Re}\brc{K_y}}{\text{A/m}}$}
\label{Fig:num:ITEM:Pappfig2y}
\psfrag{I}{$\ds\frac{\text{Re}\brc{I}}{\text{A}}$} \psfrag{x}{$x$/m}
\includegraphics[scale=0.95]{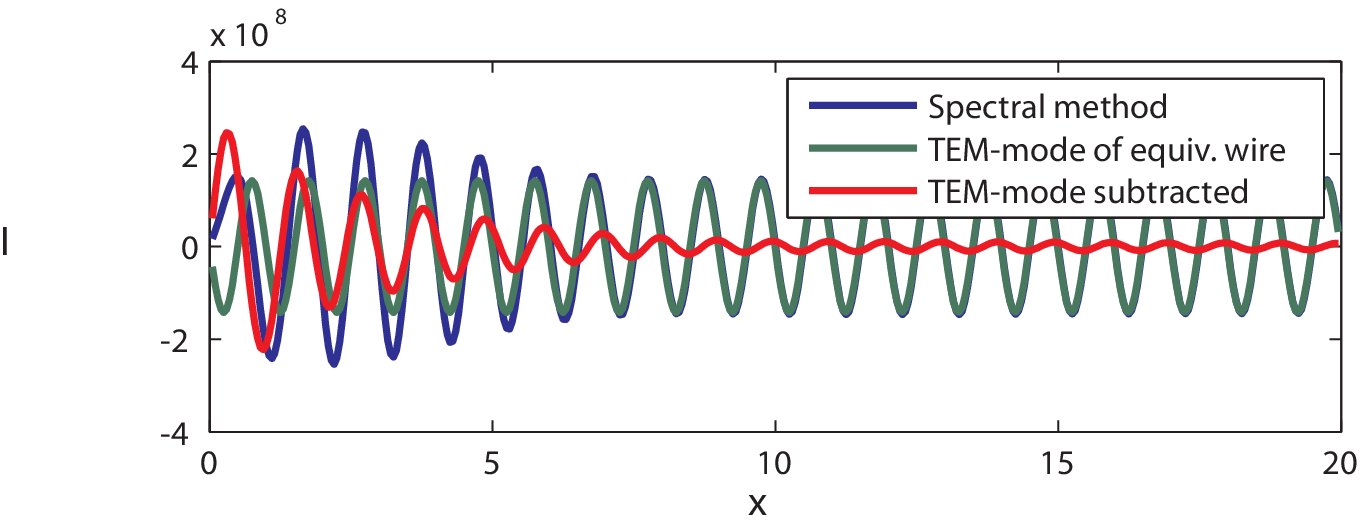} } 
\subfigure[Dipole moment $\pvv=1\uz$ Cm.]{ \psfrag{x}{$x$/m}
\psfrag{y}{$y$/m}
\psfrag{R}{$\ds\frac{\text{Re}\brc{K_x}}{\text{A/m}}$}
\psfrag{S}{$\ds\frac{\text{Re}\brc{K_y}}{\text{A/m}}$}
\label{Fig:num:ITEM:Pappfig2z}
\psfrag{I}{$\ds\frac{\text{Re}\brc{I}}{\text{A}}$} \psfrag{x}{$x$/m}
\includegraphics[scale=0.95]{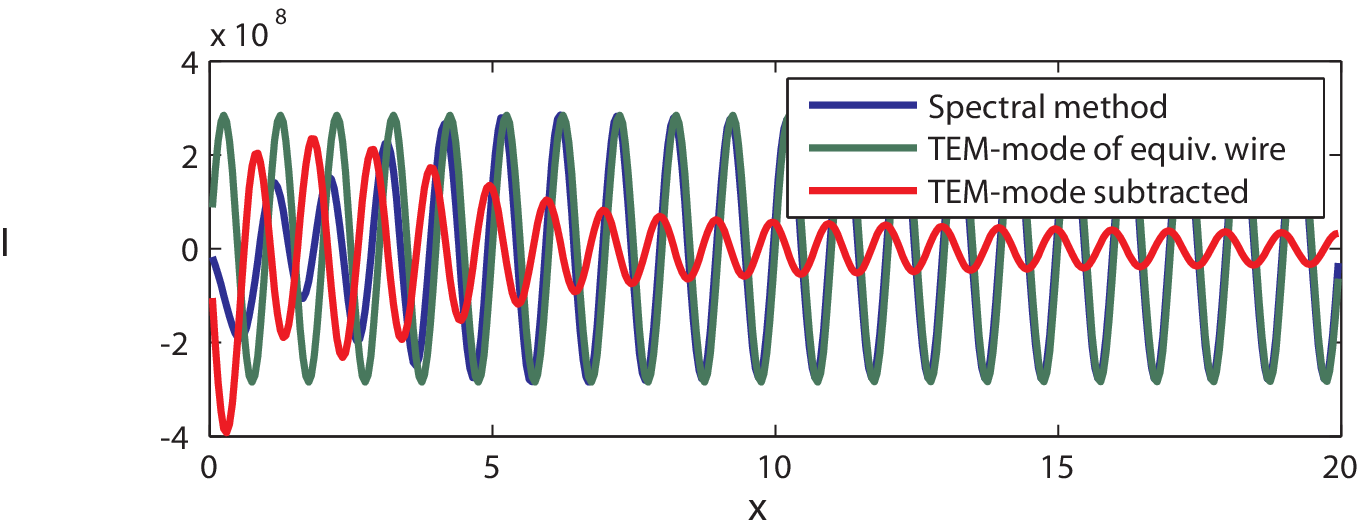} }
\caption{Strip current when $h=0.02$ m, $a=1$ m. Dipole at $\rvv_0
= 0.5\uy + 0.5\uz$ m.}
\label{Fig:num:ITEM:Pappfig2}
\end{figure}

\begin{figure}[t]
\centering
\subfigure[Dipole moment $\pvv=1\ux$ Cm.]{
\psfrag{x}{$x$/m} \psfrag{y}{$y$/m}
\psfrag{R}{$\ds\frac{\text{Re}\brc{K_x}}{\text{A/m}}$}
\psfrag{S}{$\ds\frac{\text{Re}\brc{K_y}}{\text{A/m}}$}
\label{Fig:num:ITEM:Pappfig3x}
\psfrag{I}{$\ds\frac{\text{Re}\brc{I}}{\text{A}}$} \psfrag{x}{$x$/m}
\includegraphics[scale=0.95]{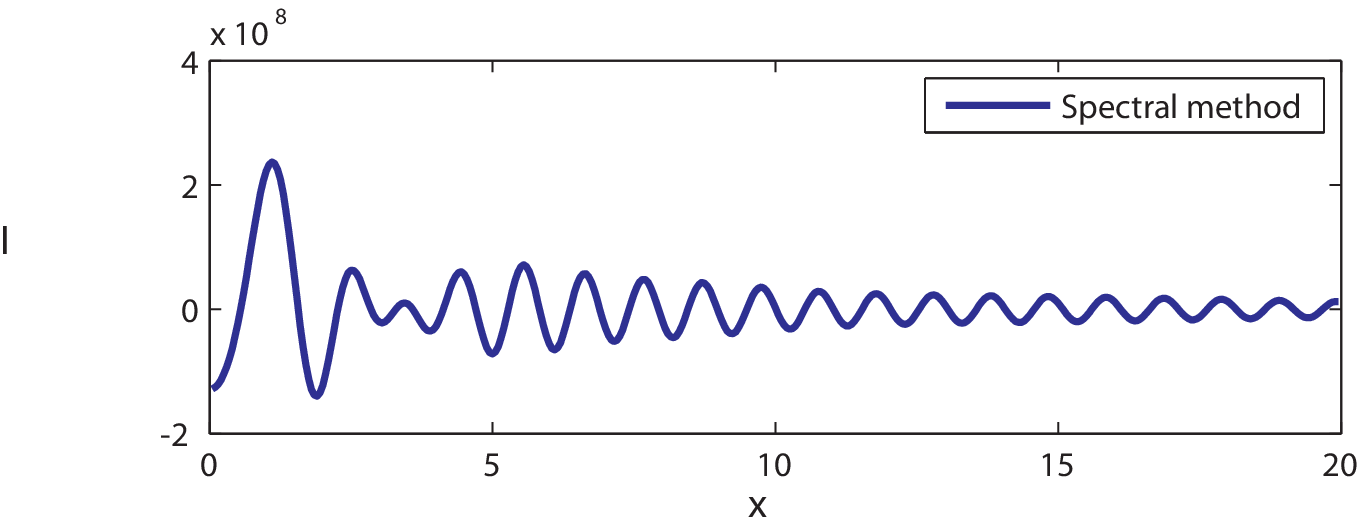} } 
\subfigure[Dipole moment $\pvv=1\uy$ Cm.]{
\psfrag{x}{$x$/m} \psfrag{y}{$y$/m}
\psfrag{R}{$\ds\frac{\text{Re}\brc{K_x}}{\text{A/m}}$}
\psfrag{S}{$\ds\frac{\text{Re}\brc{K_y}}{\text{A/m}}$}
\label{Fig:num:ITEM:Pappfig3y}
\psfrag{I}{$\ds\frac{\text{Re}\brc{I}}{\text{A}}$} \psfrag{x}{$x$/m}
\includegraphics[scale=0.95]{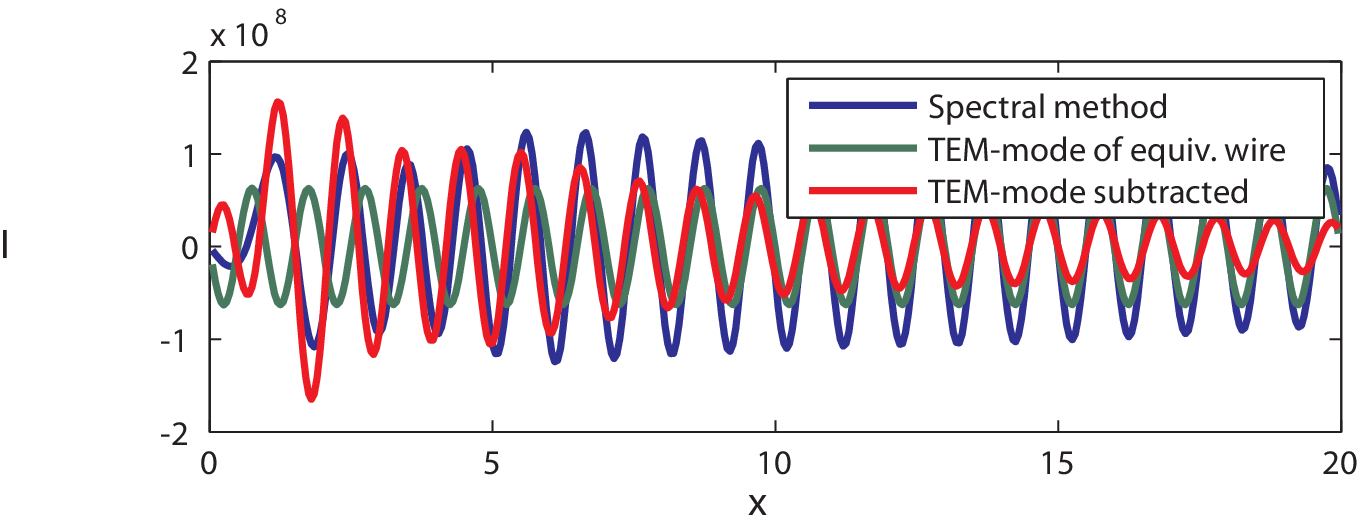} } 
\subfigure[Dipole moment $\pvv=1\uz$ Cm.]{ \psfrag{x}{$x$/m}
\psfrag{y}{$y$/m}
\psfrag{R}{$\ds\frac{\text{Re}\brc{K_x}}{\text{A/m}}$}
\psfrag{S}{$\ds\frac{\text{Re}\brc{K_y}}{\text{A/m}}$}
\label{Fig:num:ITEM:Pappfig3z}
\psfrag{I}{$\ds\frac{\text{Re}\brc{I}}{\text{A}}$} \psfrag{x}{$x$/m}
\includegraphics[scale=0.95]{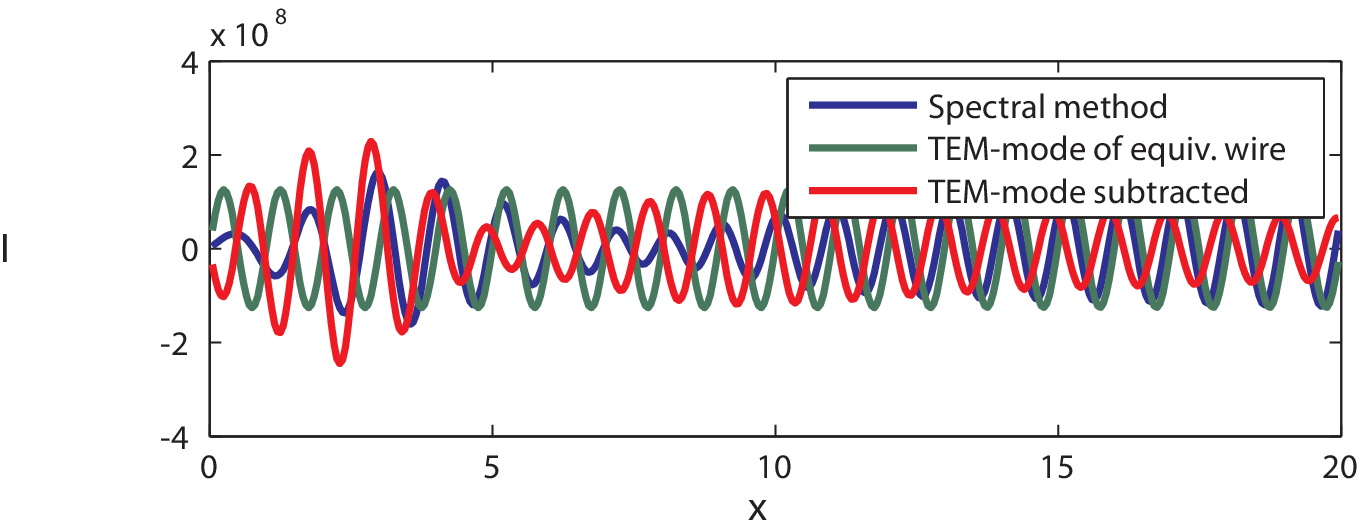} }
\caption{Strip current when $h=0.02$ m, $a=2$ m. Dipole at $\rvv_0
= \uy + \uz$ m.}
\label{Fig:num:ITEM:Pappfig3}
\end{figure}

In Figures \ref{Fig:num:ITEM:Pappfig2}-\ref{Fig:num:ITEM:Pappfig3}, we present results for the excited currents with the strip at the heights $a=1.0$ m and $a=2.0$ m, respectively (the wavelength is 1 m). In the same figures, we also present the TEM-mode current on the equivalent circular wire, as well as the difference between the total current on the strip and the TEM-mode current on the wire.

\eqref{eq:TEM:Ix} yields that  $x$-directed dipoles will not excite TEM-modes on the wire. Correspondingly, we obtain in the expression \eqref{eq:strip:c0} for the spectral current on the strip that for an $x$-directed dipole there is no pole at $k_x=k$ (the TEM-mode wavenumber), whereby the TEM-mode is not excited. Hence, the results in Figures \ref{Fig:num:ITEM:Pappfig2x} and \ref{Fig:num:ITEM:Pappfig3x} are for higher order modes on the strip only.
With $a=1.0$ m we see in Figures \ref{Fig:num:ITEM:Pappfig2y} and \ref{Fig:num:ITEM:Pappfig2z} that the excited strip current is explained reasonably by the wire TEM-mode current, apart from in the vicinity of the dipole. Roughly, the difference between the total current on the strip and the wire TEM-mode current is the contribution from the higher order modes on the strip.

Increasing further to $a=2.0$ m we see in Figures \ref{Fig:num:ITEM:Pappfig3y} and \ref{Fig:num:ITEM:Pappfig3z} that, within the distances considered, the strip current is no longer dominated by the TEM-mode.

\subsubsection{Comparison with results obtained using MoM-solver}

 Since the spectral method is designed to handle strips of infinite length, and commercial softwares that can handle infinitely long wires are not readily available, it is difficult to set up a proper comparison example. Nevertheless, we use the method of moment code NEC \cite{NEC} to compare the results for a wire of finite length with the results using the spectral method. The finite length wire is located at $\brb{x}\leq 40$ m and is left open-ended.

\begin{figure}[h]
\centering
\psfrag{R}{$\ds\frac{\text{Re}\brc{I}}{\text{A}}$} \psfrag{I}{$\ds\frac{\text{Im}\brc{I}}{\text{A}}$} \psfrag{x}{$x$/m}
\includegraphics[scale=0.95]{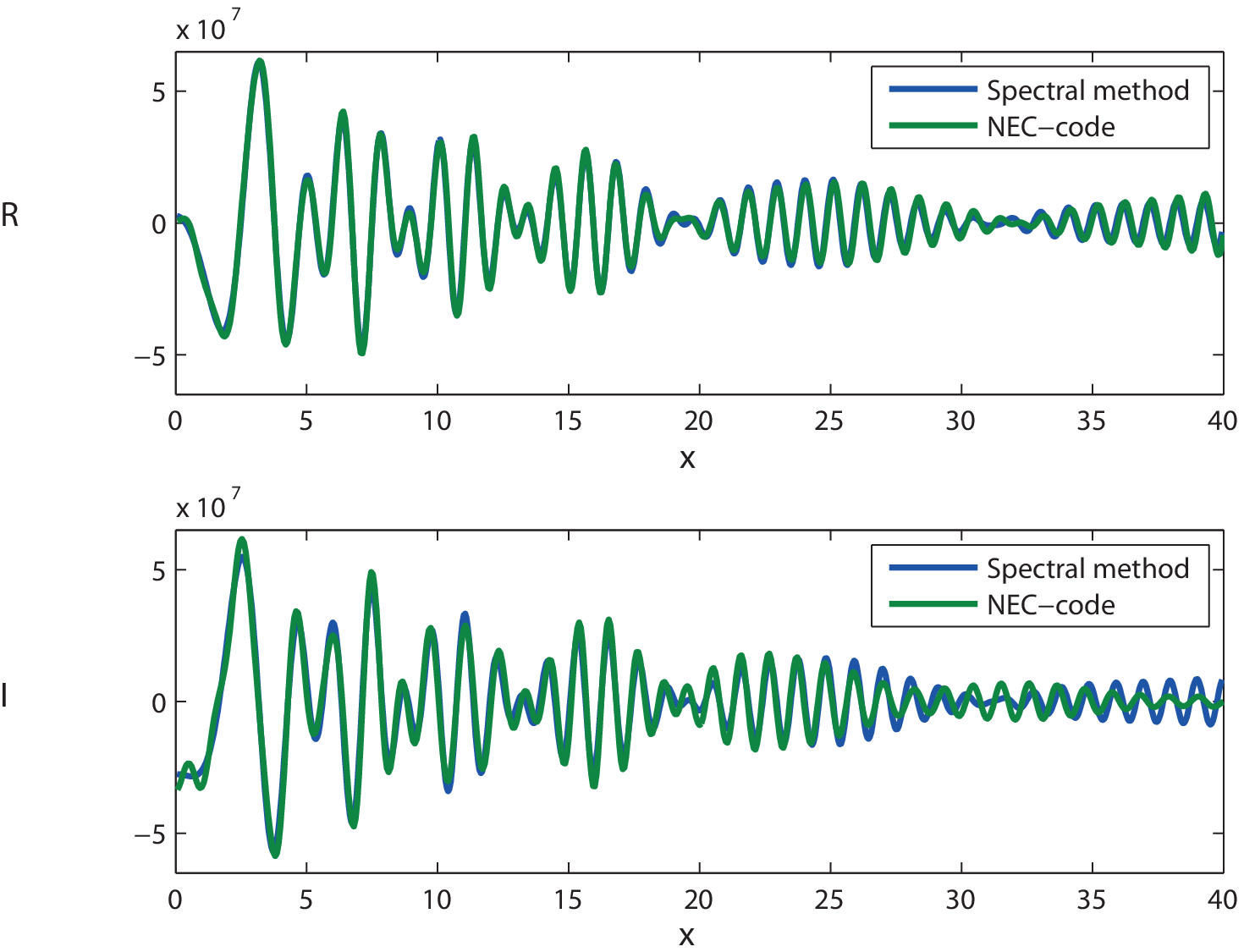}
\caption{Strip current when $h=0.02$ m, $a=6$ m. Dipole at $\rvv_0
= 6\uy + 5.5\uz$ m. Dipole moment $\pvv=1\ux$ Cm. The reference results from the NEC-code are for an open-ended wire at $\brb{x}\leq 40$ m.}
\label{Fig:num:INEC:PappfigNECx}
\end{figure}

\begin{figure}[h]
\centering
\psfrag{R}{$\ds\frac{\text{Re}\brc{I}}{\text{A}}$} \psfrag{I}{$\ds\frac{\text{Im}\brc{I}}{\text{A}}$} \psfrag{x}{$x$/m}
\includegraphics[scale=0.95]{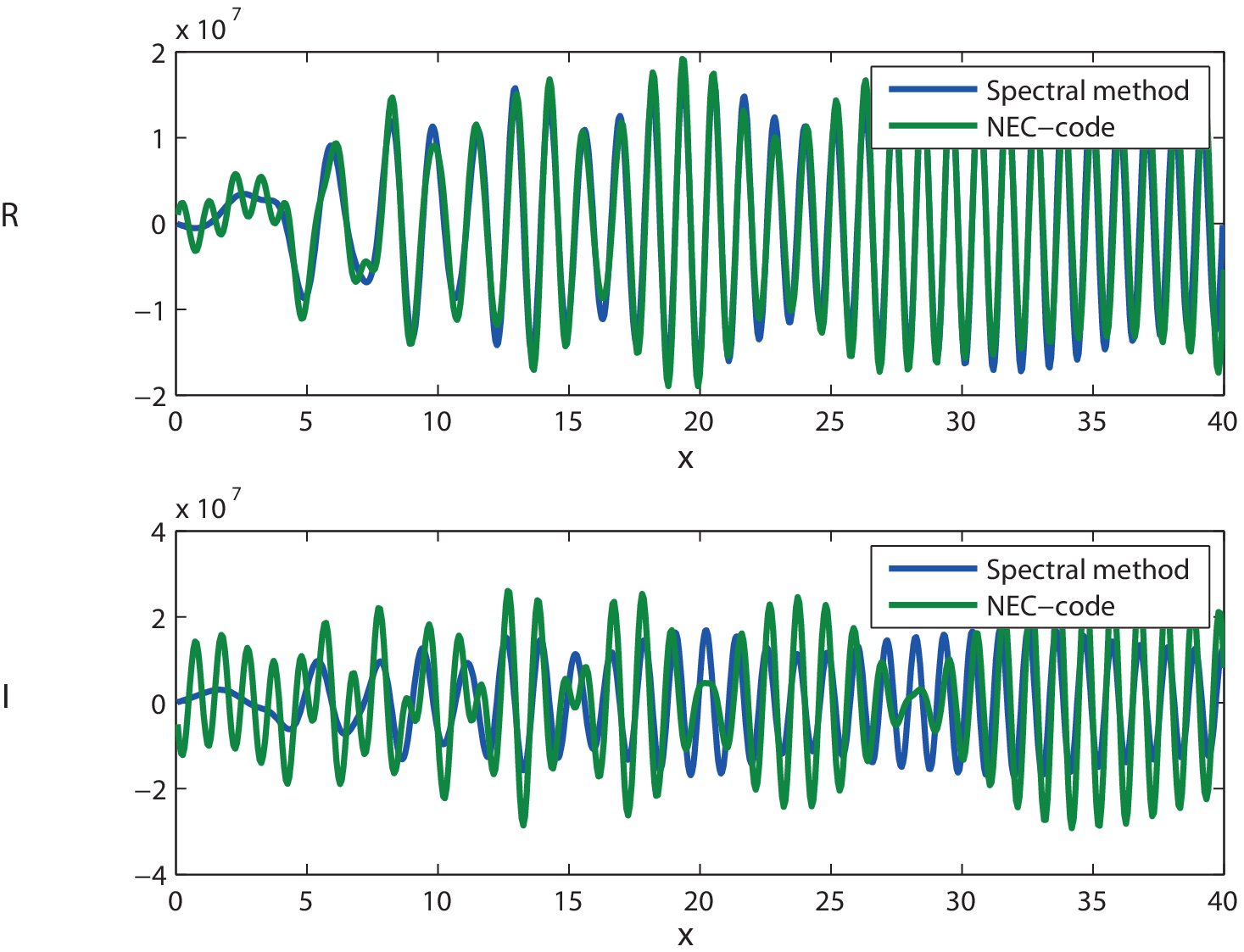}
\caption{Strip current when $h=0.02$ m, $a=6$ m. Dipole at $\rvv_0
= 6\uy + 5.5\uz$ m. Dipole moment $\pvv=1\uz$ Cm. The reference results from the NEC-code are for an open-ended wire at $\brb{x}\leq 40$ m.}
\label{Fig:num:INEC:PappfigNECz}
\end{figure}

 In Figure \ref{Fig:num:INEC:PappfigNECx} we see that with an $x$-directed dipole the agreement is quite satisfactory. The reason is the absence of the TEM-mode. Due to radiation, the higher order modes suffer from a much faster attenuation, resulting in quite small reflections from the open-ended wire in the NEC-model.  On the other hand, in Figure \ref{Fig:num:INEC:PappfigNECz} we see that with a $z$-directed dipole the agreement is far from satisfactory. Here, the TEM-mode is present, and due to its small attenuation the reflections from the open end-points of the wire creates standing waves in the NEC-model. Along the infinitely long strip there are no standing waves (but of course interference between the strip and the PEC plane).

\section{Discussion and conclusions}

\label{section:Discussion_and_conclusions}

In this paper, we have investigated high frequency wave propagation for the generic case of a metallic PEC strip above a PEC ground plane. Starting from a general formulation for a wide strip, we have verified the approximation of using only longitudinal currents on a narrow strip. Furthermore, we have verified the equivalence between narrow strips and thin wires, when proximity effects can be neglected.

Although generic, the case studied includes the key-parts regarding the mathematical analysis and the approximations made. Hence, it is in principle straight-forward to generalize to multiple parallel strips, multiple layers (including lossy media) and multiple dipole or distributed sources. However, when considering wide strips, requiring the complete system \eqref{eq:strip:cjdu1}-\eqref{eq:strip:cudj2} of equations, a generalization can be rather cumbersome. On the other hand, the simplified model for narrow strips (equivalent with thin wires), based on equation \eqref{eq:strip:c0}, is more tractable under a generalization to multiple wires and/or layers. Hence, the method outlined in this paper has potential usefulness for estimating long range propagation of high frequency waves in wire structures. Such structures can be power lines and railway feeding systems, containing both over-head wires and wires submerged into ground.


\appendix

\section{Derivation of transversal field expressions}

\label{section:app:transverse}

Here, we derive the expressions \eqref{eq:modes:Et} and
\eqref{eq:modes:Ht} for the transversal fields.

With
\begin{align}
\Evv^\pm\brn{\kvt, z}, \Hvv^\pm\brn{\kvt, z}
\propto \e^{\mp\j k_z z}
\end{align}
and  \eqref{eq:prelim:FT2} to express the fields in the source-free
(\Jvv=\vc{0}) Maxwell's equations \eqref{eq:prelim:Maxwell1} and
\eqref{eq:prelim:Maxwell2}, the spectral fields fulfil
\begin{align}
\br{\kvt \pm k_z \uz}\times\Evv^\pm & = k \eta \Hvv^\pm \label{eq:transv:1}\\
\br{\kvt \pm k_z \uz}\times\eta\Hvv^\pm & = - k \Evv^\pm
\label{eq:transv:2}
\end{align}
Decomposing \eqref{eq:transv:1}, \eqref{eq:transv:2} and the fields
into longitudinal and transversal parts, we obtain
\begin{align}
\kvt\times\Evv_\text{t}^\pm & = k\eta H_z^\pm \uz \label{eq:transv:3}\\
\kvt\times\uz E_z^\pm \pm k_z\uz\times\Evv_\text{t}^\pm & = k\eta
\Hvv_\text{t}^\pm \label{eq:transv:4}\\
\kvt\times\eta\Hvv_\text{t}^\pm & = - k E_z^\pm \uz \label{eq:transv:5}\\
\kvt\times\uz \eta H_z^\pm \pm k_z\uz\times\eta\Hvv_\text{t}^\pm & =
- k \Evv_\text{t}^\pm \label{eq:transv:6}
\end{align}
Performing the operations $\uz\times$\eqref{eq:transv:4} and
$\uz\times$\eqref{eq:transv:6}, we obtain
\begin{align}
\kvt E_z^\pm \mp k_z \Evv_\text{t} & = k \uz\times\eta\Hvv_\text{t} \label{eq:transv:7} \\
\kvt \eta H_z^\pm \mp k_z \eta \Hvv_\text{t} & = - k
\uz\times\Evv_\text{t} \label{eq:transv:8}
\end{align}
Eliminating $\uz\times\Evv_\text{t}^\pm$ from \eqref{eq:transv:4}
and \eqref{eq:transv:8}, and using \eqref{eq:prelim:kz}, we obtain
\eqref{eq:modes:Et}:
\begin{align}
\Evv_\text{t}^\pm & = \frac{-1}{\kt^2} \brh{ \pm k_z \kvt E_z^\pm -
k \uz\times\kvt \eta H_z^\pm}, \tag*{\eqref{eq:modes:Et}$^\prime$}
\end{align}
Similarly, eliminating $\uz\times\eta\Hvv_\text{t}^\pm$ from
\eqref{eq:transv:6} and \eqref{eq:transv:7}, and using
\eqref{eq:prelim:kz}, we obtain \eqref{eq:modes:Ht}:
\begin{align}
\eta \Hvv_\text{t}^\pm & = \frac{-1}{\kt^2} \brh{ \pm k_z \kvt \eta
H_z^\pm + k \uz\times\kvt E_z^\pm},
\tag*{\eqref{eq:modes:Ht}$^\prime$}
\end{align}
Note that the outlined derivation is the common way of expressing
the transversal fields in e.g. a waveguiding structure, where one
instead has $\kvt\rightarrow \j\gradT$ (see e.g. \cite{Collin2},
Section 5.1).

\newpage

\section{The spectral fields from the electric dipole}

\label{section:app:excitation}

In this appendix, we derive the expressions
\eqref{eq:AB1x}-\eqref{eq:AB1z} for the functions
$A_1\brn{\kvv_\text{t},z}$ and $B_1\brn{\kvv_\text{t},z}$, in the
spectral fields from the electric dipole.

\subsection{Spectral fields from a localised source}

\begin{figure}[h]
\centering \psfrag{m}{$\Evv, \Hvv \propto \e^{+\j k_z z}$}
\psfrag{p}{$\Evv, \Hvv \propto \e^{-\j k_z z}$} \psfrag{z}{$z$}
\psfrag{4}{$z_1$} \psfrag{3}{$z_2$} \psfrag{1}{$\mathcal{S}_1$}
\psfrag{2}{$\mathcal{S}_2$}\psfrag{J}{$\Jvv\brn{\rvv}$}
\includegraphics[scale=1.1]{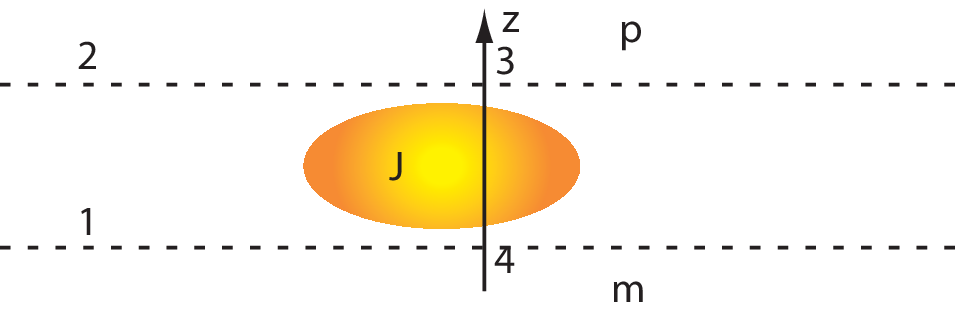}
\caption{Geometry utilized when applying the Lorentz reciprocity theorem \eqref{eq:exc:LRT}. } \label{Fig:2}
\end{figure}

First, we consider the more general case with a distributed but
localized source $\Jvv\brn{\rvv}$ located in the slab region between
the planar surfaces $\mathcal{S}_1$, at $z=z_1$, and
$\mathcal{S}_2$, at $z=z_2$ ($z_1 < z_2$); see Figure \ref{Fig:2}.
From the losses in the medium and conservation of energy, it follows
that the modes attenuate away from the slab region containing the
source.

Hence, from (\ref{eq:modes:TM_E})-(\ref{eq:modes:H_spatial}) it
holds for $z\geq z_2$  that
\begin{align}
\Evv\brn{\rvv}  = \frac{1}{4\pi^2}\int_{\mathcal{K}}\big[ &
A^+\brn{\kvv_\text{t}} \fvv^+\brn{\kvv_\text{t}} +
B^+\brn{\kvv_\text{t}} \gvv\brn{\kvv_\text{t}} \big]\e^{-\j k_z z}
\e^{-\j\vs{k}_\text{t}\cdot\vs{\rho}} \dd^2 k_\text{t} \label{eq:exc:Ep}\\[3mm]
\eta\Hvv\brn{\rvv}  = \frac{1}{4\pi^2}\int_{\mathcal{K}}\big[ &
A^+\brn{\kvv_\text{t}} \gvv\brn{\kvv_\text{t}} -
B^+\brn{\kvv_\text{t}} \fvv^+\brn{\kvv_\text{t}} \big]\e^{-\j k_z z}
\e^{-\j\vs{k}_\text{t}\cdot\vs{\rho}} \dd^2 k_\text{t}
\label{eq:exc:Hp}
\end{align}
while for $z\leq z_1$
\begin{align}
\Evv\brn{\rvv}  = \frac{1}{4\pi^2}\int_{\mathcal{K}}\big[ &
A^-\brn{\kvv_\text{t}} \fvv^-\brn{\kvv_\text{t}} +
B^-\brn{\kvv_\text{t}} \gvv\brn{\kvv_\text{t}} \big]\e^{+\j k_z z}
\e^{-\j\vs{k}_\text{t}\cdot\vs{\rho}} \dd^2 k_\text{t} \label{eq:exc:Em}\\[3mm]
\eta\Hvv\brn{\rvv}  = \frac{1}{4\pi^2}\int_{\mathcal{K}}\big[ & -
A^-\brn{\kvv_\text{t}} \gvv\brn{\kvv_\text{t}} +
B^-\brn{\kvv_\text{t}} \fvv^-\brn{\kvv_\text{t}} \big]\e^{+\j k_z z}
\e^{-\j\vs{k}_\text{t}\cdot\vs{\rho}} \dd^2 k_\text{t}
\label{eq:exc:Hm}
\end{align}

From (\ref{eq:prelim:FT1}) and (\ref{eq:prelim:FT2}), we have the
completeness relation:
\begin{align}
\frac{1}{4\pi^2}\int_{\mathcal{S}} \e^{\pm\j \br{\kvst -
\kvst^\prime} \cdot\vs{\rho}} \dS =\dirac{\kvv_\text{t}-\kvt^\prime}
\label{eq:ort:completeness}
\end{align}
where $\dirac{}$ denotes the Dirac delta distribution. From
(\ref{eq:ort:completeness}), we have the corollaries
\begin{align}
\brh{\fvv^\pm\brn{\kvv_\text{t}}\times\gvv\brn{\kvv_\text{t}^\prime}}\cdot\uz
\frac{1}{4\pi^2}\int_{\mathcal{S}}
\e^{\j\br{\vs{k}_\text{t}-\vs{k}_\text{t}^\prime} \cdot\vs{\rho}}
\dS & =\frac{k_z k}{k_\text{t}^2 \br{k_\text{t}^\prime}^2}
\kvv_\text{t}\cdot\kvv_\text{t}^\prime
\dirac{\kvv_\text{t}-\kvv_\text{t}^\prime} = \frac{k_z
k}{k_\text{t}^2}\dirac{\kvv_\text{t}-\kvv_\text{t}^\prime} \label{eq:ort:cor1}\\
\brh{\fvv^\pm\brn{\kvv_\text{t}}\times\fvv^\pm\brn{\kvv_\text{t}^\prime}}\cdot\uz
\frac{1}{4\pi^2}\int_{\mathcal{S}}
\e^{\j\br{\vs{k}_\text{t}-\vs{k}_\text{t}^\prime} \cdot\vs{\rho}}
\dS & = 0 \label{eq:ort:cor2}\\[3mm]
\brh{\fvv^\pm\brn{\kvv_\text{t}}\times\fvv^\mp\brn{\kvv_\text{t}^\prime}}\cdot\uz
\frac{1}{4\pi^2}\int_{\mathcal{S}}
\e^{\j\br{\vs{k}_\text{t}-\vs{k}_\text{t}^\prime} \cdot\vs{\rho}}
\dS & = 0 \label{eq:ort:cor3}\\[3mm]
\brh{\gvv\brn{\kvv_\text{t}}\times\gvv\brn{\kvv_\text{t}^\prime}}\cdot\uz
\frac{1}{4\pi^2}\int_{\mathcal{S}}
\e^{\j\br{\vs{k}_\text{t}-\vs{k}_\text{t}^\prime} \cdot\vs{\rho}}
\dS & = 0 \label{eq:ort:cor4}
\end{align}
The Lorentz reciprocity theorem, see e.g. \cite{Olyslager1}, yields
\begin{align}
\int_{\mathcal{S}_2}\brh{\Evv\times\Hvv^\text{test}-\Evv^\text{test}\times\Hvv}\cdot\uz\dS
-
\int_{\mathcal{S}_1}\brh{\Evv\times\Hvv^\text{test}-\Evv^\text{test}\times\Hvv}\cdot\uz\dS
= \int\Jvv\cdot\Evv^\text{test}\dV \label{eq:exc:LRT}
\end{align}
where $\brc{\Evv^\text{test}, \Hvv^\text{test}}$ is a solution to
the Maxwell's equations that is source-free in the slab region
between $\mathcal{S}_1$ and $\mathcal{S}_2$. It has been assumed
that the contribution to the total surface integral from the closing
cylindrical boundary at infinity vanishes, for which a sufficient
condition is that the source is localized and that the medium is
lossy.

Next, the mode coefficients will be calculated by inserting
(\ref{eq:exc:Ep})-(\ref{eq:exc:Hm}) into (\ref{eq:exc:LRT}) and
using the results (\ref{eq:ort:cor1})-(\ref{eq:ort:cor4}).

With $\Evv^\text{test}=\fvv^+\brn{\kvv_\text{t}}\e^{-\j k_z
z}\e^{\j\vs{k}_\text{t}\cdot\vs{\rho}},
\eta\Hvv^\text{test}=\gvv\brn{\kvv_\text{t}}\e^{-\j k_z
z}\e^{\j\vs{k}_\text{t}\cdot\vs{\rho}}$, we obtain
\begin{align}
& \int_{\mathcal{K}} \dd^2 k_\text{t}^\prime \Big[
A^+\brn{\kvt^\prime}\fvv^+\brn{\kvt^\prime}\times\gvv\brn{\kvt} +
B^+\brn{\kvt^\prime}\gvv\brn{\kvt^\prime}\times\gvv\brn{\kvt}
\nonumber \\
& - A^+\brn{\kvt^\prime}\fvv^+\brn{\kvt}\times\gvv\brn{\kvt^\prime}
+ B^+\brn{\kvt^\prime}\fvv^+\brn{\kvt}\times\fvv^+\brn{\kvt^\prime}
\Big]\cdot\uz \e^{-\j 2 k_z z_2}
 \frac{1}{4\pi^2}
\int_{\mathcal{S}}
\e^{\j\br{\vs{k}_\text{t}-\vs{k}_\text{t}^\prime}\cdot\vs{\rho}}\dS
\nonumber
\\
& - \int_{\mathcal{K}} \dd^2 k_\text{t}^\prime \Big[
A^-\brn{\kvt^\prime}\fvv^-\brn{\kvt^\prime}\times\gvv\brn{\kvt} +
B^-\brn{\kvt^\prime}\gvv\brn{\kvt^\prime}\times\gvv\brn{\kvt}
\nonumber \\
& + A^-\brn{\kvt^\prime}\fvv^+\brn{\kvt}\times\gvv\brn{\kvt^\prime}
- B^-\brn{\kvt^\prime}\fvv^+\brn{\kvt}\times\fvv^-\brn{\kvt^\prime}
\Big]\cdot\uz
 \frac{1}{4\pi^2}
\int_{\mathcal{S}}
\e^{\j\br{\vs{k}_\text{t}-\vs{k}_\text{t}^\prime}\cdot\vs{\rho}}\dS
\nonumber
\\
& = -\frac{2 k_z k}{k_\text{t}^2} A^-\brn{\kvv_\text{t}} = \eta\int
\Jvv\brn{\rvv}\cdot \fvv^+\brn{\kvv_\text{t}}\e^{-\j k_z
z}\e^{\j\vs{k}_\text{t}\cdot\vs{\rho}}\dV
\end{align}

With $\Evv^\text{test}=\fvv^-\brn{\kvv_\text{t}}\e^{+\j k_z
z}\e^{\j\vs{k}_\text{t}\cdot\vs{\rho}},
\eta\Hvv^\text{test}=-\gvv\brn{\kvv_\text{t}}\e^{+\j k_z
z}\e^{\j\vs{k}_\text{t}\cdot\vs{\rho}}$, we obtain
\begin{align}
& \int_{\mathcal{K}} \dd^2 k_\text{t}^\prime \Big[
-A^+\brn{\kvt^\prime}\fvv^+\brn{\kvt^\prime}\times\gvv\brn{\kvt} -
B^+\brn{\kvt^\prime}\gvv\brn{\kvt^\prime}\times\gvv\brn{\kvt}
\nonumber \\
& - A^+\brn{\kvt^\prime}\fvv^-\brn{\kvt}\times\gvv\brn{\kvt^\prime}
+ B^+\brn{\kvt^\prime}\fvv^-\brn{\kvt}\times\fvv^+\brn{\kvt^\prime}
\Big]\cdot\uz
 \frac{1}{4\pi^2}
\int_{\mathcal{S}}
\e^{\j\br{\vs{k}_\text{t}-\vs{k}_\text{t}^\prime}\cdot\vs{\rho}}\dS
\nonumber
\\
& - \int_{\mathcal{K}} \dd^2 k_\text{t}^\prime \Big[ -
A^-\brn{\kvt^\prime}\fvv^-\brn{\kvt^\prime}\times\gvv\brn{\kvt} -
B^-\brn{\kvt^\prime}\gvv\brn{\kvt^\prime}\times\gvv\brn{\kvt}
\nonumber \\
& + A^-\brn{\kvt^\prime}\fvv^-\brn{\kvt}\times\gvv\brn{\kvt^\prime}
- B^-\brn{\kvt^\prime}\fvv^-\brn{\kvt}\times\fvv^-\brn{\kvt^\prime}
\Big]\cdot\uz \e^{+\j 2 k_z z_1}
 \frac{1}{4\pi^2}
\int_{\mathcal{S}}
\e^{\j\br{\vs{k}_\text{t}-\vs{k}_\text{t}^\prime}\cdot\vs{\rho}}\dS
\nonumber
\\
& = -\frac{2 k_z k}{k_\text{t}^2} A^+\brn{\kvv_\text{t}} = \eta\int
\Jvv\brn{\rvv}\cdot \fvv^-\brn{\kvv_\text{t}}\e^{+\j k_z
z}\e^{\j\vs{k}_\text{t}\cdot\vs{\rho}}\dV
\end{align}

With $\Evv^\text{test}=\gvv\brn{\kvv_\text{t}}\e^{-\j k_z
z}\e^{\j\vs{k}_\text{t}\cdot\vs{\rho}},
\eta\Hvv^\text{test}=-\fvv^+\brn{\kvv_\text{t}}\e^{-\j k_z
z}\e^{\j\vs{k}_\text{t}\cdot\vs{\rho}}$, we obtain
\begin{align}
& \int_{\mathcal{K}} \dd^2 k_\text{t}^\prime \Big[
-A^+\brn{\kvt^\prime}\fvv^+\brn{\kvt^\prime}\times\fvv^+\brn{\kvt} -
B^+\brn{\kvt^\prime}\gvv\brn{\kvt^\prime}\times\fvv^+\brn{\kvt}
\nonumber \\
& - A^+\brn{\kvt^\prime}\gvv\brn{\kvt}\times\gvv\brn{\kvt^\prime} +
B^+\brn{\kvt^\prime}\gvv\brn{\kvt}\times\fvv^+\brn{\kvt^\prime}
\Big]\cdot\uz \e^{-\j 2 k_z z_2}
 \frac{1}{4\pi^2}
\int_{\mathcal{S}}
\e^{\j\br{\vs{k}_\text{t}-\vs{k}_\text{t}^\prime}\cdot\vs{\rho}}\dS
\nonumber
\\
& - \int_{\mathcal{K}} \dd^2 k_\text{t}^\prime \Big[ -
A^-\brn{\kvt^\prime}\fvv^-\brn{\kvt^\prime}\times\fvv^+\brn{\kvt} -
B^-\brn{\kvt^\prime}\gvv\brn{\kvt^\prime}\times\fvv^+\brn{\kvt}
\nonumber \\
& + A^-\brn{\kvt^\prime}\gvv\brn{\kvt}\times\gvv\brn{\kvt^\prime} -
B^-\brn{\kvt^\prime}\gvv\brn{\kvt}\times\fvv^-\brn{\kvt^\prime}
\Big]\cdot\uz
 \frac{1}{4\pi^2}
\int_{\mathcal{S}}
\e^{\j\br{\vs{k}_\text{t}-\vs{k}_\text{t}^\prime}\cdot\vs{\rho}}\dS
\nonumber
\\
& = -\frac{2 k_z k}{k_\text{t}^2} B^-\brn{\kvv_\text{t}} = \eta\int
\Jvv\brn{\rvv}\cdot \gvv\brn{\kvv_\text{t}}\e^{-\j k_z
z}\e^{\j\vs{k}_\text{t}\cdot\vs{\rho}}\dV
\end{align}

With $\Evv^\text{test}=\gvv\brn{\kvv_\text{t}}\e^{+\j k_z
z}\e^{\j\vs{k}_\text{t}\cdot\vs{\rho}},
\eta\Hvv^\text{test}=\fvv^-\brn{\kvv_\text{t}}\e^{+\j k_z
z}\e^{\j\vs{k}_\text{t}\cdot\vs{\rho}}$, we obtain
\begin{align}
& \int_{\mathcal{K}} \dd^2 k_\text{t}^\prime \Big[
A^+\brn{\kvt^\prime}\fvv^+\brn{\kvt^\prime}\times\fvv^-\brn{\kvt} +
B^+\brn{\kvt^\prime}\gvv\brn{\kvt^\prime}\times\fvv^-\brn{\kvt}
\nonumber \\
& - A^+\brn{\kvt^\prime}\gvv\brn{\kvt}\times\gvv\brn{\kvt^\prime} +
B^+\brn{\kvt^\prime}\gvv\brn{\kvt}\times\fvv^+\brn{\kvt^\prime}
\Big]\cdot\uz
 \frac{1}{4\pi^2}
\int_{\mathcal{S}}
\e^{\j\br{\vs{k}_\text{t}-\vs{k}_\text{t}^\prime}\cdot\vs{\rho}}\dS
\nonumber
\\
& - \int_{\mathcal{K}} \dd^2 k_\text{t}^\prime \Big[
A^-\brn{\kvt^\prime}\fvv^-\brn{\kvt^\prime}\times\fvv^-\brn{\kvt} +
B^-\brn{\kvt^\prime}\gvv\brn{\kvt^\prime}\times\fvv^-\brn{\kvt}
\nonumber \\
& + A^-\brn{\kvt^\prime}\gvv\brn{\kvt}\times\gvv\brn{\kvt^\prime} -
B^-\brn{\kvt^\prime}\gvv\brn{\kvt}\times\fvv^-\brn{\kvt^\prime}
\Big]\cdot\uz \e^{+\j 2 k_z z_1}
 \frac{1}{4\pi^2}
\int_{\mathcal{S}}
\e^{\j\br{\vs{k}_\text{t}-\vs{k}_\text{t}^\prime}\cdot\vs{\rho}}\dS
\nonumber
\\
& = -\frac{2 k_z k}{k_\text{t}^2} B^+\brn{\kvv_\text{t}} = \eta\int
\Jvv\brn{\rvv}\cdot \gvv\brn{\kvv_\text{t}}\e^{+\j k_z
z}\e^{\j\vs{k}_\text{t}\cdot\vs{\rho}}\dV
\end{align}
Summarizing, and using (\ref{eq:modes:fpm}) and (\ref{eq:modes:g}),
we obtain the relations
\begin{align}
A^\pm\brn{\kvv_\text{t}} & = - \frac{\eta}{2 k_z k}
\br{k_z\kvv_\text{t}\pm k_\text{t}^2\uz}\cdot\int
\Jvv\brn{\rvv}\e^{\j\br{\vs{k}_\text{t}\cdot\vs{\rho}\pm k_z
z}}\dV \label{eq:exc:Apm}\\
B^\pm\brn{\kvv_\text{t}} & = - \frac{\eta}{2 k_z} \br{\uz\times
\kvv_\text{t}} \cdot \int
\Jvv\brn{\rvv}\e^{\j\br{\vs{k}_\text{t}\cdot\vs{\rho}\pm k_z z}}\dV
\label{eq:exc:Bpm}
\end{align}


An electric dipole $\pvv$ at the location $\rvv_0=\rhov_0 + z_0\uz$
gives rise to the current density
\begin{align}
\Jvv\brn{\rvv} = \jw \pvv \dirac{\rvv-\rvv_0} \label{eq:dipoler:1}
\end{align}
where $\dirac{}$ denotes the Dirac delta distribution.

Using (\ref{eq:dipoler:1}) in (\ref{eq:exc:Apm}) and
(\ref{eq:exc:Bpm}), we obtain for the three principal orientations
of the dipole
\begin{align}
\pvv=p\ux \Rightarrow  & \begin{cases}A^\pm\brn{\kvv_\text{t}}  =
\ds - \frac{\j p k_x}{2 \ep}
\e^{\j\br{\vs{k}_\text{t}\cdot\vs{\rho}_0\pm k_z
z_0}} \\[3mm]
B^\pm\brn{\kvv_\text{t}}  =  \ds\frac{\j p k k_y}{2 \ep k_z}
\e^{\j\br{\vs{k}_\text{t}\cdot\vs{\rho}_0\pm k_z z_0}}
\end{cases} \label{eq:dipoler:2} \\[2mm]
\pvv=p\uy \Rightarrow  & \begin{cases}A^\pm\brn{\kvv_\text{t}}  =
\ds - \frac{\j p k_y}{2 \ep}
\e^{\j\br{\vs{k}_\text{t}\cdot\vs{\rho}_0\pm k_z
z_0}} \\[3mm]
B^\pm\brn{\kvv_\text{t}}  =  \ds -\frac{\j p k k_x}{2 \ep k_z}
\e^{\j\br{\vs{k}_\text{t}\cdot\vs{\rho}_0\pm k_z z_0}}
\end{cases} \label{eq:dipoler:3} \\[2mm]
\pvv=p\uz \Rightarrow  & \begin{cases}A^\pm\brn{\kvv_\text{t}}  =
\ds \pm \frac{\j p k_\text{t}^2}{2 \ep k_z}
\e^{\j\br{\vs{k}_\text{t}\cdot\vs{\rho}_0\pm k_z
z_0}} \\[3mm]
B^\pm\brn{\kvv_\text{t}}  =  0
\end{cases} \label{eq:dipoler:4}
\end{align}

Finally, with
\begin{align}
A_1\brn{\kvv_\text{t},z} = A^\pm\brn{\kvv_\text{t}}\e^{\mp\j k_z z},
\qquad  B_1\brn{\kvv_\text{t},z} = B^\pm\brn{\kvv_\text{t}}\e^{\mp\j
k_z z}
\end{align}
we obtain the expressions \eqref{eq:AB1x}-\eqref{eq:AB1z}.

\section{Numerical evaluation of spectral integrals}

\label{section:app:integrals}

Here, we describe the numerical evaluation of the spectral integrals
in (\ref{eq:strip:cjdu1})-(\ref{eq:strip:cudj2}).

Since the integrals run over the semi-infinite interval $0 < k_y <
\infty$ and are not open for easy analytic evaluation, they must be
approximated over a finite interval $0 < k_y < k_{y,\text{max}}$. To
increase the rate of convergence with increasing $k_{y,\text{max}}$,
we identify and subtract off leading behaviours of the integrands
for large $k_y$ that can be integrated analytically.

First, the longitudinal wavenumber is
\begin{align}
k_z = \sqrt{k^2 - k_x^2 - k_y^2} \label{app:int:kz}
\end{align}
With $\text{Im}\brc{k_z} < 0$ and $0 < k_y < \infty$, it holds for
$k_y \gg \brb{k_x}, \brb{k}$ that $k_z \approx -\j k_y$.

Hence, for large $k_y$ we have in the left-hand sides of
(\ref{eq:strip:cjdu1})-(\ref{eq:strip:cudj2}) integrands that behave
as
\begin{align}
& \frac{1-\e^{-\j 2 k_z a}}{k_z}\bJ{\nu_1}{k_y h}\bJ{\nu_2}{k_y h}
\approx \frac{1-\e^{-2 k_y a}}{-\j
 k_y}\bJ{\nu_1}{k_y h}\bJ{\nu_2}{k_y h} \approx \frac{\j}{k_y}\bJ{\nu_1}{k_y h}\bJ{\nu_2}{k_y h} \\
& \frac{\br{1-\e^{-\j 2 k_z a}}\br{k^2-k_y^2}}{k_y^2
k_z}\bJ{\nu_1}{k_y h}\bJ{\nu_2}{k_y h} \approx \frac{-\j}{k_y}
\bJ{\nu_1}{k_y h}\bJ{\nu_2}{k_y h}
\end{align}
From \cite{GradshteynRyzhik}, formula 6.574, we have the following
results for non-negative integers $m$ and $n$:
\begin{align}
\int_0^\infty \frac{\bJ{2m}{k_y h}\bJ{2n}{k_y h}}{k_y}\dd k_y & =
\frac{\updelta_{mn}}{2\br{m+n}}, \quad \br{m,n}\neq \br{0,0}
\label{app:int:mnj}
\\[3mm]
\int_0^\infty \frac{\bJ{2m+1}{k_y h}\bJ{2n+1}{k_y h}}{k_y}\dd k_y &
= \frac{\updelta_{mn}}{2\br{m+n+1}} \label{app:int:mnu}
\end{align}
Using (\ref{app:int:mnj}), the integrals in (\ref{eq:strip:cjdu1})
and (\ref{eq:strip:cjdu2}) are approximated as
\begin{align}
& \int_0^\infty \frac{1-\e^{-\j 2 k_z a}}{k_z} \bJ{2m}{k_y
h}\bJ{2n}{k_y h}\dd k_y \nonumber \\ & \approx
\int_0^{k_{y,\text{max}}} \brh{ \frac{1-\e^{-\j 2 k_z a}}{k_z}
-\frac{\j}{k_y}} \bJ{2m}{k_y
h}\bJ{2n}{k_y h}\dd k_y + \frac{\j \:\updelta_{mn}}{2\br{m+n}}, \quad \br{m,n}\neq \br{0,0}  \\[4mm]
& \int_0^\infty \frac{\br{1-\e^{-\j 2 k_z a}}\br{k^2-k_y^2}}{k_y^2
k_z} \bJ{2m}{k_y h}\bJ{2n}{k_y h}\dd k_y \nonumber \\ & \approx
\int_0^{k_{y,\text{max}}} \brh{ \frac{\br{1-\e^{-\j 2 k_z
a}}\br{k^2-k_y^2}}{k_y^2 k_z} + \frac{\j}{k_y}} \bJ{2m}{k_y
h}\bJ{2n}{k_y h}\dd k_y - \frac{\j \:\updelta_{mn}}{2\br{m+n}}
\end{align}
For the case $m=n=0$, the first integral in (\ref{eq:strip:cjdu1})
is approximated as
\begin{align}
\int_0^\infty \frac{1-\e^{-\j 2 k_z a}}{k_z} & \text{J}_0^2\brn{k_y
h} \dd k_y  \approx \int_0^{1/h} \frac{1-\e^{-\j 2 k_z
a}}{k_z}\text{J}_0^2\brn{k_y h} \dd k_y \nonumber \\
 & + \int_{1/h}^{k_{y,\text{max}}} \brh{ \frac{1-\e^{-\j 2 k_z
a}}{k_z} -\frac{\j}{k_y}}\text{J}_0^2\brn{k_y h} \dd k_y +
\int_1^\infty \frac{\text{J}_0^2\brn{u}}{u}\dd u,
\end{align}
where $\ds \int_1^\infty \frac{\text{J}_0^2\brn{u}}{u}\dd u \approx
0.3438831082$, evaluated as a generalized hypergeometric function.
Using (\ref{app:int:mnu}), the integrals in (\ref{eq:strip:cudj1})
and (\ref{eq:strip:cudj2}) are approximated as
\begin{align}
& \int_0^\infty \frac{1-\e^{-\j 2 k_z a}}{k_z} \bJ{2m+1}{k_y
h}\bJ{2n+1}{k_y h}\dd k_y \nonumber \\ & \approx
\int_0^{k_{y,\text{max}}} \brh{ \frac{1-\e^{-\j 2 k_z a}}{k_z}
-\frac{\j}{k_y}} \bJ{2m+1}{k_y
h}\bJ{2n+1}{k_y h}\dd k_y + \frac{\j \:\updelta_{mn}}{2\br{m+n+1}}  \\[4mm]
& \int_0^\infty \frac{\br{1-\e^{-\j 2 k_z a}}\br{k^2-k_y^2}}{k_y^2
k_z} \bJ{2m+1}{k_y h}\bJ{2n+1}{k_y h}\dd k_y \nonumber \\ & \approx
\int_0^{k_{y,\text{max}}} \brh{ \frac{\br{1-\e^{-\j 2 k_z
a}}\br{k^2-k_y^2}}{k_y^2 k_z} + \frac{\j}{k_y}} \bJ{2m+1}{k_y
h}\bJ{2n+1}{k_y h}\dd k_y - \frac{\j \:\updelta_{mn}}{2\br{m+n+1}}
\end{align}
In the right hand sides of
(\ref{eq:strip:cjdu1})-(\ref{eq:strip:cudj1}), the exponential
functions typically guaranty rapid convergence of the integrals,
except for small values of $a-z_0$, i.e. when the dipole source is
close to the plane containing the strip. To account for such cases,
we for $x$-directed dipoles extract the following explicit integral
for the leading behaviours (\cite{GradshteynRyzhik} 6.621):
\begin{align}
\int_0^\infty \frac{\e^{-k_y b}}{k_y}\bJ{m}{k_y h} \dd k_y =
\frac{1}{m}\br{\frac{h}{2b}}^m \text{F}\brn{\frac{m}{2},
\frac{m+1}{2}; m+1; -\br{\frac{h}{b}}^2}, \;\; m = 1, 2, \ldots
\label{app:int:rhsx}
\end{align}
where $\text{F}\brn{}$ is the hypergeometric function. For $y$- and
$z$-directed dipoles we extract the following explicit integral for
the leading behaviours (\cite{GradshteynRyzhik} 6.611):
\begin{align}
\int_0^\infty \e^{-k_y b}\bJ{m}{k_y h} \dd k_y =
\frac{\brh{\sqrt{b^2+h^2}-b}^m}{h^m\sqrt{b^2 + h^2}}, \;\; m = 0, 1,
\ldots \label{app:int:rhsyz}
\end{align}
In both \eqref{app:int:rhsx} and \eqref{app:int:rhsyz}, the complex
variable $b$ has $\text{Re}\brc{b}=b\mp z_0$ and
$\text{Im}\brc{b}=\pm y_0$.

\begin{figure}[h]
\centering \psfrag{m}{$-k$} \psfrag{p}{$k$}
\psfrag{1}{$\text{Im}\brc{k_x}$} \psfrag{2}{$\text{Re}\brc{k_x}$}
 \psfrag{3}{$-k_{x,\text{max}}$}  \psfrag{4}{$k_{x,\text{max}}$}
  \psfrag{d}{$\Delta k_x$}   \psfrag{t}{$-\Delta k_x$}
\includegraphics[scale=1]{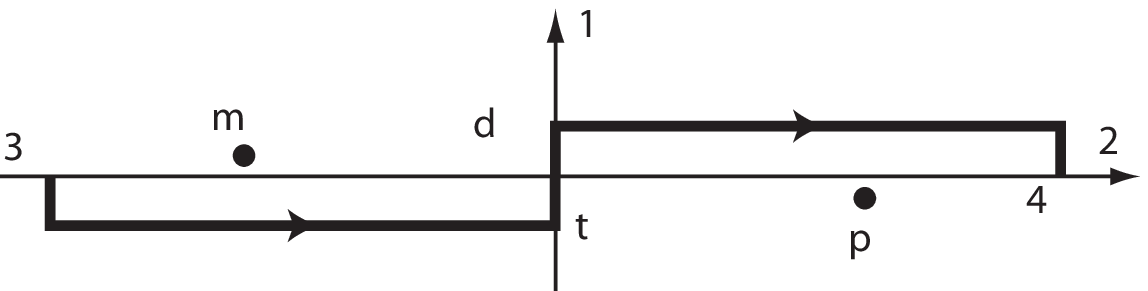}
\caption{Integration contour in the numerical approximation of
\eqref{eq:strip:cd:spat}.} \label{Fig:Ikx}
\end{figure}

The integrals in (\ref{eq:strip:cjdu1})-(\ref{eq:strip:cudj2}) have
been computed numerically by the routine \cte{quadl} in the Matlab
software. We notice that the integrands have poles where $k_z=0$. By
considering the medium as generally lossy, the constitutive
parameters have non-zero imaginary parts, moving the poles off the
real axis. However, in the near lossless limit these poles are close
to real axis, causing numerical problems in the integrals. To
alleviate this problem, the numerical approximations of the
integrals in \eqref{eq:strip:cd:spat} are calculated by deforming
the integration path to the contour shown in Figure \ref{Fig:Ikx}.

With $\Delta k_x \gg -\text{Im}\brc{k}$, we obtain better numerical
stability for the integrals in
(\ref{eq:strip:cjdu1})-(\ref{eq:strip:cudj2}), and for the
subsequent integrals in \eqref{eq:strip:cd:spat}. The upper limit on
$\Delta k_x$ is in practice determined by the acceptable exponential
growth in the term $\e^{\j k_x x}$ (see \eqref{eq:strip:cd:spat}).

\section{The TEM-mode current along an $x$-directed circular wire}

\label{section:app:TEMmod}

In this appendix, we derive the expressions for the TEM-mode current on a circular wire, used in Section \ref{section:TEMcomp} for comparison purposes.

Consider a PEC circular wire oriented in the $x$-direction. The wire
has radius $s$ and is centered at $\br{y=0,z=a}$ above the PEC
ground plane ($s < a$). Note that in the present context, $y$ and
$z$ are the transversal directions, subscripted with t, while $x$ is
the longitudinal direction.

Applying \eqref{eq:exc:LRT} in the present context
, the TEM-mode excitation coefficients become
\begin{align}
D_0^\pm = - \frac{\ds\int_\mathcal{V}\Jvv\cdot\Evv_\text{t}\e^{\pm\j
kx}\dV}{2\ds\int_\mathcal{A}\br{\Evv_\text{t}\times\Hvv_\text{t}}\cdot\ux\dA}
\end{align}
where $\mathcal{V}$ is the volume containing the source and
$\mathcal{A}$ is the halfplane $\br{z>0} \cap \br{-\infty < y <
\infty}$ expect for the cross-section of the wire.
$\Evv_\text{t}\brn{y,z}$ and $\Hvv_\text{t}\brn{y,z}$ are the modal field patterns in the cross-sectional plane.

From the TEM-mode relation
$\Hvv_\text{t}=\eta^{-1}\ux\times\Evv_\text{t}$ (where
$\ux\cdot\Evv_\text{t}=0$), we obtain
\begin{align}
D_0^\pm = -
\frac{\eta}{2}\frac{\ds\int_\mathcal{V}\Jvv\cdot\Evv_\text{t}\e^{\pm\j
kx}\dV}{\ds\int_\mathcal{A}E_\text{t}^2\dA} \label{eq:TEM:exc2}
\end{align}
Introducing, the scalar potential function $\Phi\brn{y,z}$,
$\Evv_\text{t} = -\gradT\Phi$ and $\divT\Evv_\text{t}=\lapT\Phi=0$
together with the Gauss' theorem in the plane yields
\begin{align}
\int_\mathcal{A}E_\text{t}^2\dA =
\int_\mathcal{A}\brb{\gradT\Phi}^2\dA =
\int_\mathcal{A}\brh{\divT\br{\Phi\gradT\Phi}-\Phi\lapT\Phi}\dA =
\oint_\mathcal{C} \Phi\un\cdot\gradT\Phi \dl
\end{align}
where $\mathcal{C}=\mathcal{C}_0 + \mathcal{C}_\infty +
\mathcal{C}_s$, in which $\mathcal{C}_0$ is the line $z=0$ that is
closed with the curve $\mathcal{C}_\infty$ at infinity (with $z
>0$), and $\mathcal{C}_s$ is the circumference of the wire.
With $\Phi=0$ on $\mathcal{C}_0$ and $\mathcal{C}_\infty$, with the
constant value $\Phi=\Phi_s$ on $\mathcal{C}_s$, and the boundary
condition $\sigma = \ep\un\gradT\Phi$ for the surface charge
density, we obtain
\begin{align}
\int_\mathcal{A} E_\text{t}^2\dA = \Phi_s\oint_{\mathcal{C}_s}
\frac{\sigma}{\ep} \dl = \Phi_s\frac{\lambda}{\ep}
\label{eq:TEM:exc4}
\end{align}
where $\lambda$ is the line charge density on the wire.

By standard image theory \cite{Cheng}, we obtain
\begin{align}
\Phi\brn{y,z} =
\frac{\lambda}{2\pi\ep}\ln\sqrt{\frac{y^2+z^2+a^2-s^2+2z\sqrt{a^2-s^2}}{y^2+z^2+a^2-s^2-2z\sqrt{a^2-s^2}}}
\label{eq:TEM:Phi}
\end{align}
On the wire, i.e. when $y^2+\br{z-a}^2 = s^2$, \eqref{eq:TEM:Phi}
yields
\begin{align}
\Phi_s = \frac{\lambda}{2\pi\ep} \ln\frac{a+\sqrt{a^2-s^2}}{s} =
\frac{\lambda}{2\pi\ep}\text{arccosh}\brn{\frac{a}{s}}
\label{eq:TEM:Phis}
\end{align}
From \eqref{eq:TEM:Phi}, we obtain
\begin{align}
\Evv_\text{t}\brn{y,z}=-\gradT\Phi\brn{y,z} =
\frac{\lambda}{2\pi\ep}\evv\brn{y,z} \label{eq:TEM:Et}
\end{align}
where
\begin{align}
\evv\brn{y,z} = \frac{y\uy +\br{z-\sqrt{a^2-s^2}}\uz}{y^2
+\br{z-\sqrt{a^2-s^2}}^2} - \frac{y\uy
+\br{z+\sqrt{a^2-s^2}}\uz}{y^2
+\br{z+\sqrt{a^2-s^2}}^2}\label{eq:TEM:e}
\end{align}
With the dipole $\pvv$ at $\rvv_0$, yielding $\Jvv\brn{\rvv}=\jw
\pvv\dirac{\rvv-\rvv_0}$, it follows from \eqref{eq:TEM:exc2},
\eqref{eq:TEM:exc4}, \eqref{eq:TEM:Phis} and \eqref{eq:TEM:Et} that

\begin{align}
D_0^\pm = - \frac{\eta\ep \jw
\pvv\cdot\evv\brn{y_0,z_0}}{2\lambda\text{arccosh}\brn{a/s}}\e^{\pm
\j k x_0} \label{eq:TEM:D0pm}
\end{align}
Hence, the $\Hvv$-field becomes
\begin{align}
\Hvv & = \pm D_0^\pm \text{H}\brn{x-x_0}\e^{\mp\j k
x}\Hvv_\text{t}\brn{y,z} \nonumber \\
&  = - \text{sgn}\brn{x-x_0} \frac{\eta\ep \jw
\pvv\cdot\evv\brn{y_0,z_0}}{2\lambda\text{arccosh}\brn{a/s}}\e^{\mp
\j k \brb{x-x_0}} \Hvv_\text{t}\brn{y,z}
\end{align}
To determine the current distribution along the wire, we must find
the modal current associated with
\begin{align}
\Hvv_\text{t}\brn{y,z} &
=\frac{1}{\eta}\ux\times\Evv_\text{t}\brn{y,z}\nonumber
\\
& = \frac{\lambda}{2\pi\ep\eta}\brh{ \frac{y\uz
-\br{z-\sqrt{a^2-s^2}}\uy}{y^2 +\br{z-\sqrt{a^2-s^2}}^2} -
\frac{y\uz -\br{z+\sqrt{a^2-s^2}}\uy}{y^2 +\br{z+\sqrt{a^2-s^2}}^2}
}
\end{align}
On the ground plane, the surface current density becomes
\begin{align}
\Kvv\brn{y} = \uz\times\Hvv_\text{t}\brn{y,z=0} = -
\frac{\lambda}{\pi\ep\eta}\cdot\frac{\sqrt{a^2-s^2}}{y^2 + a^2 -
s^2}\ux
\end{align}
which yields that the oppositely directed current on the wire
becomes
\begin{align}
I_s = - \int_{-\infty}^\infty \Kvv\brn{y}\cdot\ux\dd y  = \frac{2
\lambda}{\pi\ep\eta} \int_0^\infty \frac{\sqrt{a^2-s^2}}{y^2 + a^2 -
s^2}\dd y = \frac{\lambda}{\ep\eta}
\end{align}
Hence, the TEM-mode current distribution on the wire becomes
\begin{align}
I_\text{TEM}\brn{x} & = - \text{sgn}\brn{x-x_0} \frac{\eta\ep \jw
\pvv\cdot\evv\brn{y_0,z_0}}{2\lambda\text{arccosh}\brn{a/s}}\e^{\mp
\j k \brb{x-x_0}} I_s \nonumber \\
& = - \text{sgn}\brn{x-x_0} \frac{\jw
\pvv\cdot\evv\brn{y_0,z_0}}{2\text{arccosh}\brn{a/s}}\e^{\mp \j k
\brb{x-x_0}} \label{eq:TEM:I_TEM}
\end{align}
Explicitly, for the three principal directions of the dipole,
it follows from \eqref{eq:TEM:e} and \eqref{eq:TEM:I_TEM} that
\begin{align}
\pvv = p\ux \Rightarrow & \; I_\text{TEM}\brn{x}=0 \label{eq:TEM:Ix}\\[5mm]
\pvv = p\uy \Rightarrow & \nonumber \\
I_\text{TEM}\brn{x}  = - & \frac{\jw p \;
\text{sgn}\brn{x-x_0}}{2\text{arccosh}\brn{a/s}} \brh{
\frac{y_0}{y_0^2 +\br{z_0-\sqrt{a^2-s^2}}^2} - \frac{y_0 }{y^2
+\br{z_0+\sqrt{a^2-s^2}}^2} }\label{eq:TEM:Iy}\\[5mm]
\pvv = p\uz \Rightarrow & \nonumber \\
I_\text{TEM}\brn{x}   = - & \frac{\jw p \;
\text{sgn}\brn{x-x_0}}{2\text{arccosh}\brn{a/s}} \brh{
\frac{z_0-\sqrt{a^2-s^2}}{y_0^2 +\br{z_0-\sqrt{a^2-s^2}}^2} -
\frac{z_0+\sqrt{a^2-s^2}}{y_0^2 +\br{z_0+\sqrt{a^2-s^2}}^2} }
\label{eq:TEM:Iz}
\end{align}
The expressions \eqref{eq:TEM:Iy} and \eqref{eq:TEM:Iz} hold for any
wire radius $0 < s < a$, but if $s\ll a, \brb{a-z_0}$, we can use
$\sqrt{a^2-s^2}\approx a$ and
$\text{arccosh}\brn{a/s}\approx\ln\brn{2 a/s}$, cf.
\eqref{eq:TEM:Phis}.


\begin{thebibliography}{99}

%
%



\bibitem{CozzaDemoulin} A. Cozza and B. D\'emoulin,
\cte{On the Modeling of Electric Railway Lines for the Assessment of
Infrastructure Impact in Radiated Emission Tests of Rolling Stock},
{\em IEEE Trans. Electromagn. Compat.}, vol. 50, no. 3, pp. 566-576,
Aug. 2008.




\bibitem{AmirshahiKavedrah} P. Amirshahi and M. Kavedrah,
\cte{High-Frequency Characteristics of Overhead Multiconductor Power
Lines for Broadband Communications}, {\em IEEE J. Sel. Areas in
Communic.}, vol. 24, no. 7, pp. 1292-1303, July 2006.


\bibitem{SeeETAL} K. Y. See, P. L. So, A. Kamarul, and E. Gunawan,
\cte{Radio-Frequency Common-Mode Noise Propagation Model for
Power-Line Cable}, {\em IEEE Trans. Power Deliv.}, vol. 20, no. 4,
pp. 2443-2449, Oct. 2005.


\bibitem{LazaropoulosCottis} A. G. Lazaropoulos and P. G. Cottis,
\cte{Transmission Characteristics of Overhead Medium-Voltage Power
Line Communication Channels}, {\em IEEE Trans. Power Deliv.}, vol.
24, no. 3, pp. 1164-1173, July 2009.


\bibitem{SartenaerDelogne} T. Sartenaer and P. Delogne,
\cte{Deterministic Modeling of the (Shielded) Outdoor Power Line
Channel Based on the Multiconductor Transmission Line Equations},
{\em IEEE J. Sel. Areas in Communic.}, vol. 24, no. 7, pp.
1277-1291, July 2006.



\bibitem{PoljakETAL1} D. Poljak, S. Antonijevic, K. E. K. Drissi, and
K. Kerroum, \cte{Transient Response of Straight Thin Wires Located
at Different Heights Above a Ground Plane Using Antenna Theory and
Transmission Line Approach}, {\em IEEE Trans. Electromagn. Compat.},
vol. 52, no. 1, pp. 108-116, Feb. 2010.

\bibitem{AlyonesETAL} S. Alyones, C. W. Bruce, and A. K. Buin,
\cte{Numerical Methods for Solving the Problem of Electromagnetic
Scattering by a Thin Finite Conducting Wire}, {\em IEEE Trans.
Antennas Propag.}, vol. 55, no. 6, pp. 1856-1861, June 2007.




\bibitem{PoljakETAL2} D. Poljak, V. Doric, F. Rachidi, K. E. K. Driss,
K. Kerroum, S. V. Tkachenko, and S. Sesnic, \cte{Generalized Form of
Telegapher's Equations for the Electromagnetic Field Coupling to
Buried Wires of Finite Length}, {\em IEEE Trans. Electromagn.
Compat.}, vol. 51, no. 2, pp. 331-337, May. 2009.


\bibitem{TheethayiETAL1} N. Theethayi, Y. Baba, F. Rachidi, and R. Thottappillil,
\cte{On the Choice Between Transmission Line Equations and Full-Wave
Maxwell´s Equations for Transient Analyis of Buried Wires}, {\em
IEEE Trans. Electromagn. Compat.}, vol. 51, no. 2, pp. 331-337, May.
2009.


\bibitem{PetracheETAL} E. Petrache, F. Rachidi, M. Paolone, C. A. Nucci,
V. A. Rakov, and M. A. Uman, \cte{Lightning Induced Distrurbances in
Buried Cables - Part I: Theory}, {\em IEEE Tr. Electrom. Compat.},
vol. 47, no. 3, pp. 498-508, Aug. 2005.

\bibitem{TheethayiETAL2} N. Theethayi, R. Thottappillil, M. Paolone, C. A. Nucci, and
F. Rachidi, \cte{External Impedance and Admittance of Buried
Horizontal Wires for Transient Studies Using Transmission Line
Analysis}, {\em IEEE Tr. Diel. Insul.}, vol. 14, no. 3, pp. 751-761,
June 2007.




\bibitem{NeffReed} H. P. Neff and D. A. Reed,
\cte{The Effect of Secondary Scattering on the Induced Current in a
Long Wire Over an Imperfect Ground from an Incident EMP}, {\em IEEE
Tr. Ant. Prop.}, vol. 37, no. 12, pp. 1554-1558, Dec. 1989.

\bibitem{Tesche} F. M. Tesche,
\cte{Comparison of the Transmission Line and Scattering Models for
Computing the HEMP Response of Overhead Cables}, {\em IEEE Tr.
Electrom. Compat.}, vol. 34, no. 2, pp. 93-99, May 1992.

\bibitem{BridgesShafai} G. E. J. Bridges and L. Shafai,
\cte{Plane Wave Coupling to Multiple Conductor Transmission Lines
Above a Lossy Earth}, {\em IEEE Tr. Electrom. Compat.}, vol. 31, no.
1, pp. 21-33, Feb. 1989.




\bibitem{Butler} C. M. Butler,
\cte{General Solutions of the Narrow Strip (and Slot) Integral
Equations}, {\em IEEE Tr. Ant. Prop.}, vol. 33, no. 10, pp.
1085-1090, Oct. 1985.

\bibitem{Tsalamengas} J. L. Tsalamengas, J. G. Fikioris, and B. T. Babili,
\cte{Direct and efficient solutions of integral equations for
scattering from strips and slots}, {\em J. Appl. Phys}, vol. 66, no.
1, pp. 69-80, July 1989.

\bibitem{Shively} D. Shively,
\cte{Scattering from Perfectly Conducting and Resistive Strips on a
Grounded Dielectric Slab}, {\em IEEE Tr. Ant. Prop.}, vol. 42, no.
4, pp. 552-556, Apr. 1994.

\bibitem{BaccarelliETAL} P. Baccarelli, P. Burghignoli, G. Lovat,
S. Paulotto, F. Mesa, and D. R. Jackson, \cte{Direct model
transition from space wave to surface wave leakage on microstrip
lines}, {\em Radio Science}, vol. 40, RS6017,
doi:10.1029/2005RS003286, 2005.

\bibitem{LovatETAL} G. Lovat, P. Burghignoli, F. Capolino, D. R. Jackson, and D. R. Wilton,
\cte{High-gain Omnidirectional Radiation Patterns from a Metal Strip
Grating Leaky-Wave Antenna}, {\em IEEE Ant. Prop. Soc., AP-S
Int.Symp.}, pp. 5797-5800, 2007.

\bibitem{KaganovskyShavit} Y. Kaganovsky and R. Shavit,
\cte{Analysis of Radiation From a Line Source in a Grounded
Dielectric Slab Covered by a Metal Strip Grating}, {\em IEEE Tr.
Ant. Prop.}, vol. 57, no. 1, pp. 135-143, Jan. 2009.





\bibitem{Norgren2003} M. Norgren, \cte{A simple approach to quasi-TEM
analysis of a planar multiconductor structure embedded in an
elliptically stratified environment}, {\em Microwave Opt. Technol.
Lett.}, vol. 36, no. 1, pp 20-24, Jan. 2003.

\bibitem{Lindell1981} I. V. Lindell, \cte{On the Quasi-TEM Modes in
Inhomogeneous Multiconductor Transmission Lines}, {\em IEEE
Transactions on Microwave Theory and Techniques}, vol. 29, no. 8, pp
812-817, Aug. 1981.


\bibitem{LindellGu1987} I. V. Lindell, \cte{Theory of Time-Domain Quasi-TEM
Modes in Inhomogeneous Multiconductor Lines}, {\em IEEE Transactions
on Microwave Theory and Techniques}, vol. 35, no. 10, pp 839-907,
Oct. 1987.


\bibitem{GradshteynRyzhik} I. S. Gradshteyn and I. M. Ryzhik,
{\em Table of Integrals, Series and Products}, Academic Press, New
York, (1980).



\bibitem{RachidiTkachenko} F. Rachidi and S. V. Tkachenko, {\em
Electromagnetic Field Interaction with Transmission Lines: From
Classical Theory to HF Radiation Effects}, Advances in Electrical
Engineering and Electromagnetics, Series Volume 5, WIT Press, 2008.

\bibitem{Collin2} R. E. Collin,
{\em Field Theory of Guided Waves}, 2nd ed., IEEE Press, New York,
1991.


\bibitem{Midya} S. Midya,
{\em Conducted and Radiated Electromagnetic Interference in Modern
Electrified Railways with Emphasis on Pantograph Arcing}, PhD
thesis, Royal Institute of Technology, Stockholm, 2009.


\bibitem{FelsenMarkuwitz1994} L. B. Felsen and N. Markuwitz,
{\em Radiation and Scattering of Waves}, EEE Press, New York, 1994

\bibitem{Hallen} E. Hall\'en,
{\em Electromagnetic Theory}, Chapman \& Hall, London, 1962.

\bibitem{LindellETAL1994} I. V. Lindell, A. H. Sihvola, S. A.
Tretyakov, and A. J. Viitanen, {Electromagnetic Waves in Chiral and
Bi-Isotropic Media}, Artech House, London, 1994.



\bibitem{NEC} http://en.wikipedia.org/wiki/Numerical\_Electromagnetics\_Code

\bibitem{Olyslager1} F. Olyslager,
{\em Electromagnetic Waveguides and Transmission Lines}, Oxford
University Press, 1991.

\bibitem{VanBladel} J. Van Bladel,
{\em Singular Electromagnetic Fields and Sources}, IEEE Press, New
York, 1995.

\bibitem{Cheng} D. K. Cheng,
{\em  Field and Wave Electromagnetics}, Addison Wesley, 1989.


\end{thebibliography}
\end{document}